\documentclass[12pt, draftclsnofoot, onecolumn, twoside]{IEEEtran}
\usepackage{amssymb,amsmath,mathrsfs,bm,graphicx,color,cite,subfigure,multirow,cases}
\usepackage[]{hyperref}

\allowdisplaybreaks

\newtheorem{lemma}{Lemma}
\newtheorem{theorem}{Theorem}
\newtheorem{corollary}{Corollary}

\newcommand{\blue}{\color{black}}

\newcommand{\by}{ {\bf y} }
\newcommand{\bu}{ {\bf u} }
\newcommand{\bv}{ {\bf v} }

\newcommand{\bi}{ {\bf i} }
\newcommand{\bn}{ {\bf n} }

\newcommand{\pardef}{\stackrel{\Delta}{=}}
\newcommand{\e}{ \mathrm{e} }
\newcommand{\diff}{ \mathrm{d} }

\begin{document}
\title{{\blue  Energy Harvesting Noncoherent Cooperative Communications}}
\author{Peng Liu,  
        Saeed Gazor, 
        Il-Min Kim, 
        and Dong In Kim
%
\thanks{P. Liu is with the Department of Electrical Engineering, Stanford University, Stanford, CA, 94305, USA (e-mail: pengliu1@stanford.edu).}
\thanks{S. Gazor and I.-M. Kim are with the Department of Electrical and Computer Engineering, Queen's University, Kingston, Ontario, K7L 3N6, Canada (e-mail: gazor@queensu.ca; ilmin.kim@queensu.ca).}
%
\thanks{D. I. Kim is with the school of Information and Communication Engineering, Sungkyunkwan University (SKKU), Suwon, Korea (e-mail: dikim@skku.ac.kr)}
}



\maketitle

\vspace{-1.8cm}

\begin{abstract}
This paper investigates simultaneous wireless information and power transfer (SWIPT) in energy harvesting (EH) relay systems. Unlike existing SWIPT schemes requiring the channel state information (CSI) for \emph{coherent} information delivery, we propose a \emph{noncoherent} SWIPT framework for decode-and-forward (DF) relay systems bypassing the need for CSI and consequently saving energy in the network. The proposed SWIPT framework embraces \emph{power-splitting noncoherent DF} (PS-NcDF) and \emph{time-switching noncoherent DF} (TS-NcDF) in a unified form, and supports arbitrary $M$-ary noncoherent frequency-shift keying (FSK) and differential phase-shift keying (DPSK). Exact (noncoherent) maximum-likelihood detectors (MLDs) for PS-NcDF and TS-NcDF are developed in a unified form, which involves integral evaluations yet serves as the optimum performance benchmark for noncoherent SWIPT. To reduce the computational cost of the exact MLDs, we also propose \emph{closed-form} approximate MLDs achieving near-optimum performance, thus serving as a \emph{practical} solution for noncoherent SWIPT. Numerical results demonstrate a performance tradeoff between the first and second hops through the adjustment of time switching or power splitting parameters, whose optimal values minimizing the symbol-error rate (SER) are strictly between 0 and 1. We demonstrate that $M$-FSK results in a significant energy saving over $M$-DPSK for $M\geq 8$; thus $M$-FSK may be more suitable for EH relay systems.

\end{abstract}


\section{Introduction}  \label{sec:intro}
The wireless information transfer (WIT) and wireless power transfer (WPT) technologies share the common origin of the electromagnetic waves, which is a direct consequence of the fact that radio frequency (RF) signals carry energy as well as information at the same time. Nevertheless, prior research activities and industrial developments on WIT have been carried out independently from those on WPT, due to {\blue the fact that WIT and WPT operate with rather different receiver power sensitivities, i.e., $-60$ dBm for information receivers and $-10$ dBm for energy receivers  \cite{X.Lu.accepted.Kim,X.Zhou2013.11}. As a result, WIT and WPT have distinct receiver architectures and design objectives. In particular, WIT makes use of the \emph{active} mixer for maximal information delivery rate \cite{A.Goldsmith2005book}, whereas WPT utilizes the \emph{passive} rectenna for maximal energy transmission efficiency \cite{N.Shinohara2014book}.} It is not until recently that the dual use of RF signals for \emph{simultaneous wireless information and power transfer} (SWIPT) \cite{L.R.Varshney2008.7,P.Grover2010.6} has been studied, giving rise to an appealing solution for energy-constrained wireless systems such as cooperative relay networks.

\subsection{Related Works}
Recent advances on SWIPT have demonstrated conflicting objectives in designing WIT for maximal information delivery rate and WPT for maximal energy transmission efficiency, as witnessed by the fundamental \emph{rate-energy tradeoff} \cite{L.R.Varshney2008.7}. Varshney first introduced the notion of the  rate-energy region for SWIPT in additive white Gaussian noise (AWGN) channels \cite{L.R.Varshney2008.7}, which was then generalized to frequency-selective fading channels \cite{P.Grover2010.6}. These pioneering studies, however, were based on an ideal receiver circuit which simultaneously decodes information and harvests energy from the same signal without any loss \cite{L.R.Varshney2008.7,P.Grover2010.6}. It has been recognized that this ideal receiver architecture \cite{L.R.Varshney2008.7,P.Grover2010.6} may be difficult to implement given the current state-of-the-art of electronic circuits. A practical receiver architecture incorporating both energy harvesting (EH) and information decoding (ID) circuits into a single terminal was proposed in \cite{R.Zhang2013.5}. Based on this practical architecture, the wireless terminal may operate either in the time switching mode \cite{L.Liu2013.1}, in which the receiver perform EH and ID in a time-division fashion, or in the power splitting mode \cite{L.Liu2013.9}, in which the receiver splits the received signal into two streams for separate EH and ID at the same time. The corresponding rate-energy tradeoff was studied for single-input single-output (SISO) \cite{X.Zhou2013.11, L.Liu2013.9} and multiple-input multiple-output (MIMO) channels \cite{R.Zhang2013.5}. The power allocation, beamforming design, and optimization schemes were studied to achieve various tradeoffs between EH and ID. The throughput maximization for a multi-user system was addressed in \cite{H.Ju2014.1}. The optimum power allocation maximizing the information rate while guaranteeing that the harvested energy is above a threshold was studied for broadband systems where perfect instantaneous channel state information (CSI) was available at the transmitter \cite{K.Huang2013.12}. Moreover, beamforming scheme maximizing the harvested energy subject to information rate constraint was studied under imperfect CSI in \cite{Z.Xiang2012.8}. Similarly, beamforming design maximizing the information rate subject to EH constraint was addressed in \cite{C.Xing2013.4}. The precoder maximizing the energy efficiency in MIMO wiretap channel subject to secrecy rate and power constraints was studied in \cite{H.Zhang2014.9}. These works \cite{L.R.Varshney2008.7,P.Grover2010.6,R.Zhang2013.5,L.Liu2013.1,L.Liu2013.9,X.Zhou2013.11,K.Huang2013.12,Z.Xiang2012.8,C.Xing2013.4,H.Ju2014.1,H.Zhang2014.9} represent typical applications of SWIPT in point-to-point (P2P) communications.

Recently, the SWIPT has also found its application in energy-constrained wireless relay networks. In EH relay systems, the source and destination typically have dedicated power supplies from battery or power grid, while the relays are powered by the RF signals radiated by the source. The throughput performance of an amplify-and-forward (AF) relay network employing EH relays was studied for both delay-limited and delay-tolerant applications in \cite{A.A.Nasir2013.7}. The benefits of EH relays and the differences between EH relay networks and conventional relay networks were investigated in \cite{B.Medepally2010.11}. In \cite{I.Ahmed2013.4}, a joint power allocation and relay selection scheme for AF relay networks with EH ability was studied.  A greedy switching strategy between EH and data relaying was investigated for a dual-hop AF network, and the outage probability was analyzed in \cite{I.Krikidis2012.11}. Besides AF relay systems, SWIPT has also been studied for decode-and-forward (DF) relay systems. Specifically, a DF network consisting of multiple source-destination pairs and one EH relay was studied in \cite{Z.Ding2013.12}, where the outage probability was studied taking into account the spatial randomness of the source-destination pairs. For the same system, the diversity performance and power allocation were studied in \cite{Z.Ding2014.2}. Similarly, a DF EH network consisting of one or multiple sources, multiple relays, and one destination was studied in \cite{Z.Ding2014.8}, and a more general DF system consisting of a multiple source-destination pairs and multiple EH relays in the presence of random interferences was studied in \cite{I.Krikidis2014.3}, where the small-scale fading and the large-scale path-loss were taken into account in the outage probability analysis.

\subsection{Motivation}
Due to the far-field isotropic RF radiation, the energy transfer efficiency of WPT (e.g., microwave power transfer) via RF signals is very low. Beamforming techniques (based on large antenna arrays) can substantially boost the energy efficiency by forming focused energy beams using the instantaneous CSI. The existing works \cite{L.R.Varshney2008.7,P.Grover2010.6,R.Zhang2013.5,L.Liu2013.1,L.Liu2013.9,X.Zhou2013.11,K.Huang2013.12,Z.Xiang2012.8,C.Xing2013.4,H.Ju2014.1,
H.Zhang2014.9, A.A.Nasir2013.7, B.Medepally2010.11, I.Ahmed2013.4, I.Krikidis2012.11, Z.Ding2013.12,Z.Ding2014.2,Z.Ding2014.8,I.Krikidis2014.3} on SWIPT have either explicitly or implicitly assumed \emph{coherent} WIT which requires the instantaneous CSI for information decoding. {\blue For EH relay systems featuring coherent WIT, the source needs to periodically send training symbols for channel estimation, which results in an increased network signaling overhead. Moreover,  sophisticated channel estimation algorithm needs to be implemented in a \emph{distributed} manner among spatially separated wireless terminals, which inevitably increases the hardware complexity and results in extra energy consumption. The consumed energy due to channel estimation and training becomes a serious issue especially for EH relay systems, as the consumed power reduces the net harvested energy for future data relaying. Therefore, \emph{coherent} SWIPT requiring channel estimation might be a costly approach for energy-constrained applications. As a remedy, the \emph{noncoherent} SWIPT which eliminates the need for training signalings and the channel estimation, and consequently, consumes much less energy, might be an  energy-efficient and cost-effective solution for EH relay systems.} To the best of our knowledge, however, the \emph{noncoherent} SWIPT has not been studied in the literature, which motivated this work.

The existing works on SWIPT have been devoted to the \emph{information-theoretic} performance analysis in terms of the rate-energy tradeoff, ergodic capacity, outage probability, etc. However, insights into the SWIPT regarding the bit or symbol error rate performance associated with an actual detection scheme from the \emph{communication-theoretic} perspective are not available. To the best of our knowledge, the maximum-likelihood detector (MLD) achieving noncoherent SWIPT has not been studied yet. In this paper, we will tackle the noncoherent MLDs which characterize the optimum performance benchmark for ID in EH relay systems.

\subsection{Contribution}
The main objective of this paper is to develop the \emph{noncoherent} SWIPT framework and practical MLDs for EH relay networks. We consider an EH relay system consisting of a source, a destination, and multiple EH relays, which can harvest energy directly from the RF signals radiated by the source and utilize the harvested energy to decode-and-forward the source information to the destination. The main contributions of the paper are summarized as follows:

\begin{itemize}
  \item A unified noncoherent SWIPT framework embracing both \emph{power-splitting noncoherent DF} (PS-NcDF) and \emph{time-switching noncoherent DF} (TS-NcDF) protocols is developed, which supports any $M$-DPSK or $M$-FSK signalings enabling the noncoherent SWIPT.
  \item Following the unified noncoherent SWIPT framework, we derive the exact MLDs for PS-NcDF and TS-NcDF in a unified form, for both $M$-FSK and $M$-DPSK signalings. The exact MLDs involve numerical integral evaluations; yet are useful for characterizing the optimum performance benchmark for ID in noncoherent SWIPT.
  \item To reduce the computational cost of the exact MLDs, we develop \emph{closed-form} approximate MLDs with substantially lower complexity. Numerical results demonstrate that the proposed approximate MLDs achieve almost identical performance to the exact MLDs, thus constituting a practical solution for \emph{noncoherent}  SWIPT.
  \item Useful insights into the noncoherent SWIPT are obtained from the communication-theoretic perspective in terms of the symbol-error rate (SER). Specifically, we demonstrate a performance tradeoff between the first and second hops through the adjustment of the time switching coefficient or power splitting factor. Moreover, the optimal values of these system parameters corresponding to the minimum symbol-error rate (SER) are strictly between 0 and 1. Finally, we demonstrate that $M$-FSK is an energy-efficient solution for noncoherent SWIPT, especially when $M\geq 8$.
  \item As a byproduct of this work, we obtain generic analytical results dealing with the distributions of circularly symmetric complex Gaussian (CSCG) random variables and their transformations. These generic results may be of general usefulness for other different applications. Moreover, an \emph{exact} expression of the transition (error) probability is developed for $M$-DPSK. In addition, a tight closed-form approximation of the transition probability is developed. To the best of our knowledge, these analytical results have not been reported in the literature.
\end{itemize}

\subsection{Organization and Notation}

Section \ref{sec:sysmodel} introduces the system model of EH relay networks. Section \ref{sec:NcDF_relaying} develops the noncoherent SWIPT framework for EH relay networks. Section \ref{sec:math} presents the mathematical fundamentals developed in this paper. Section \ref{sec:MLD} proposes the closed-form approximate as well as the exact MLDs. Section \ref{sec:numerical} evaluates the performance of the proposed detectors and Section \ref{sec:con} concludes the paper.

\emph{Notation}: We use $A \pardef B$ to denote that $A$ is defined by $B$. Also, $(\cdot)^*$, $(\cdot)^T$, $(\cdot)^H$,  $\|\cdot\|$, $\mathbb{E}(\cdot)$, and $\ln(\cdot)$  denote the conjugate, transpose, conjugate transpose, 2-norm, expectation, and natural logarithm, respectively. $\bm{I}_n$ denotes an $n\times n$ identity matrix, $\bm{0}$ denotes an all-zero column vector, $\bi_{n}$ denotes a column vector with 1 at its $n$-th entry and 0 elsewhere, and $\measuredangle z$ denotes the phase of $z$ over a $2\pi$ interval of interest.  Finally, $\bm{x}\sim\mathcal{CN}(\bm{\mu},\bm{\Sigma})$ indicates that $\bm{x}$ is a CSCG random vector with mean $\bm{\mu}$ and covariance $\bm{\Sigma}$.


\section{System Model}\label{sec:sysmodel}


Consider an EH relay system comprising a source terminal $\mathrm{T}_0$, a destination terminal $\mathrm{T}_d$, and $K$ relays $\mathrm{T}_r$, $r=1,2,\cdots,K$. The source and destination are powered by dedicated energy sources such as battery or power grid, while the relays have no power supply and instead can harvest energy from the RF signals radiated by the source. Using the harvested energy, the relays assist the source-destination communication through the noncoherent DF relaying, enabling SWIPT in EH relay systems where the instantaneous CSI is not needed.

We consider a composite fading model where the wireless channels are subject to both small-scale fading and large-scale path-loss. Note that the path-loss can lead to a serious deterioration in the power transfer efficiency and thus cannot be neglected in EH systems \cite{I.Krikidis2014.3}. Let $\mathcal{L}_{ij}$ denote the path-loss associated with the channel from the transmit terminal $i$ to the receive terminal $j$, and $h_{ij}\sim\mathcal{CN}(0,1)$ denotes the corresponding small-scale Rayleigh fading coefficient, for $ij\in\{0d,0r,rd\}_{r=1}^K$. Note that this channel modeling implies that the actual variances of the fading coefficients $h_{ij}$ are absorbed into the corresponding path-loss components $\mathcal{L}_{ij}$, and thus, a general \emph{asymmetric} fading model is essentially considered by choosing distinct path-loss components $\mathcal{L}_{ij}$ for different wireless links. We consider the \emph{noncoherent} communication where the instantaneous CSIs, $h_{ij}$, $ij\in\{0d,0r,rd\}_{r=1}^K$, are unknown to all terminals in the network. The SWIPT is accomplished through the noncoherent signalings such as $M$-FSK and $M$-DPSK. Specifically, the source broadcasts RF signals according to the noncoherent $M$-FSK or $M$-DPSK modulation. The relays utilize the harvested energy to \emph{noncoherently} decode and forward the information to the destination, where the ultimate \emph{noncoherent} detection is performed. Thus, the end-to-end information transfer is accomplished through pure noncoherent signalings.

Due to the absence of the CSI, sophisticated protocols such as distributed space-time coding, beamforming, and relay selection which require the CSI cannot be applied. Instead, we adopt the time-division based protocol which coordinates multiple terminals' transmissions in different time intervals. Specifically, while the source transmits its signal, all relays stay silent and harvest energy from the source RF signals. After the source transmission, the relays decode the source information and forward it to the destination in different time intervals, allowing the destination to combine multiple received signals and thus enabling the diversity gain. Note that the time-division relaying protocol allows the developed detectors to be readily integrated into existing time-division multiple access (TDMA) systems, and it has been considered in numerous works including EH relay systems \cite{A.A.Nasir2013.7, B.Medepally2010.11, I.Krikidis2012.11, Z.Ding2013.12,Z.Ding2014.2,Z.Ding2014.8,I.Krikidis2014.3} and self-powered relay networks \cite{C.Xing2013.3,C.Xing2012.9,P.Liu2013.9}.

{\blue For ease of the exposition, we focus on the single-antenna transmission over frequency-flat fading channels in this paper. However, the extension to the multi-antenna and/or frequency-selective fading is possible. Specifically, by applying the multi-carrier transmission such as orthogonal
frequency division multiplexing (OFDM), the frequency-selective fading is automatically turned into flat fading, thus enabling the application of our scheme. Moreover, the extension to the multiple-input multiple-output (MIMO) configuration over each hop is also possible. The remaining issue is to design the noncoherent/differential space-time codes specifically suitable for EH relay systems, which may be considered as a future work.}


\section{A Unified Noncoherent SWIPT Framework}\label{sec:NcDF_relaying}
{\blue For EH systems, the main motivation to use noncoherent modulation schemes such as noncoherent FSK and DPSK\footnote{{\blue We do not consider $M$-QAM modulation in this paper because it is typically used for coherent communication where the instantaneous CSI is available.}} coupled with appropriate noncoherent detection schemes is that the periodic training symbols and sophisticated channel estimation algorithms are completely eliminated, which results in substantial reduction of the signaling overhead and processing burden, and consequently reduces the energy consumption in the network.} In this section, by modifying the power splitting and time switching receiver structures for \emph{coherent} SWIPT \cite{X.Zhou2013.11} where the instantaneous CSI is required, we develop a unified \emph{noncoherent} SWIPT framework embracing both PS-NcDF and TS-NcDF requiring no instantaneous CSI.

\subsection{PS-NcDF Relaying}\label{sub:PS_NcDF}
\begin{figure}[!t]%
\centering
\subfigure[][PS-NcDF]
{\includegraphics[width=0.4\columnwidth]{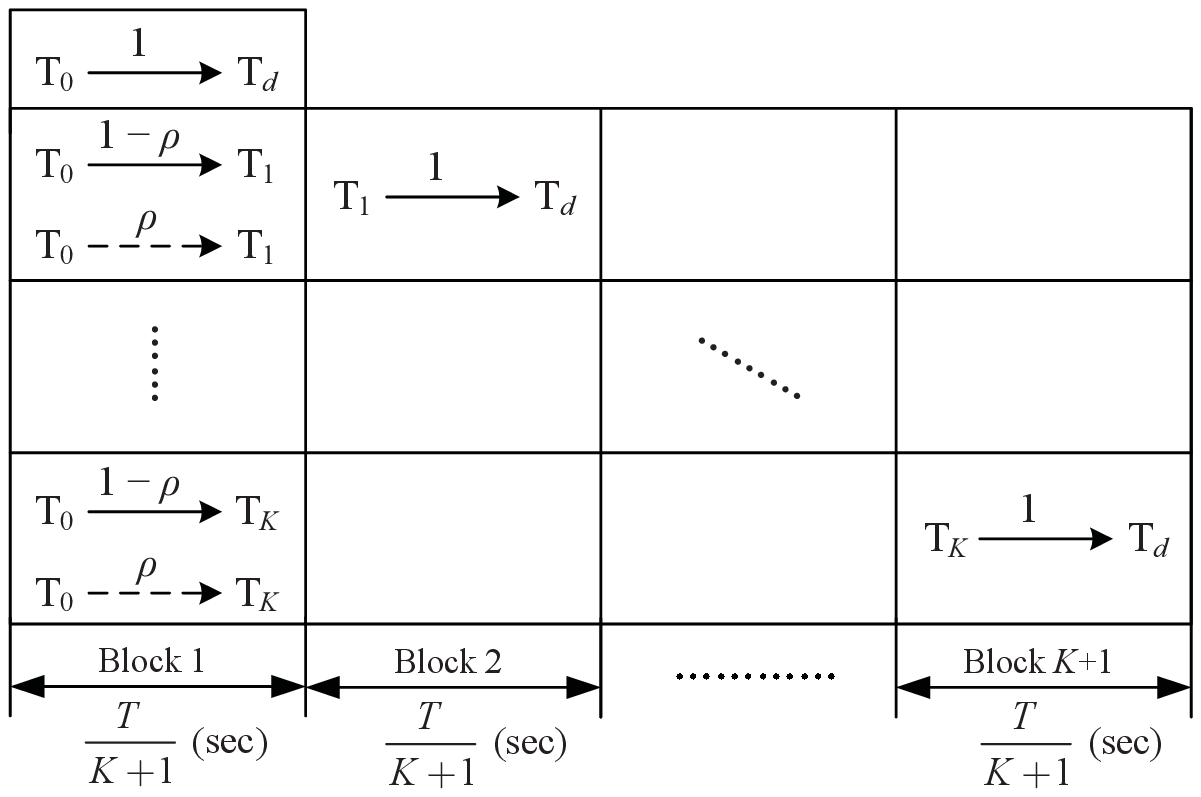}\label{fig:PsNcDF}}
\hfil
\subfigure[][TS-NcDF]
{\includegraphics[width=0.4\columnwidth]{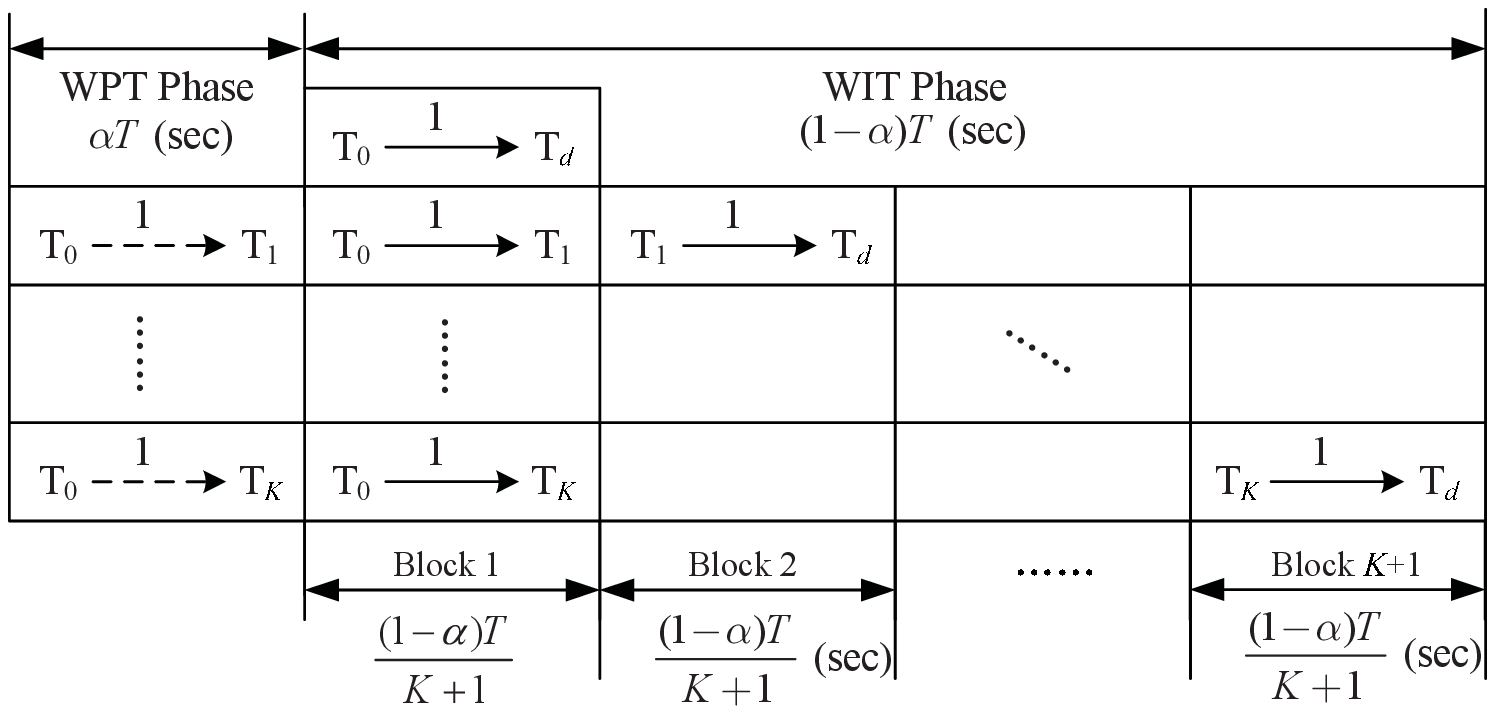}\label{fig:TsNcDF}}
\caption{PS-NcDF and TS-NcDF protocols, where the solid and dashed arrow lines represent the information flow and energy flow, respectively. The numbers above the arrow lines indicate the portions of the power utilized for corresponding purposes.}
\label{fig:PS_TS_NcDF}
\end{figure}

{\blue In the PS-NcDF protocol, the WPT and WIT circuits at each relay terminal function simultaneously while the power fed to each circuit is a portion split from the front-end RF received signal.} Let $T$ (sec) denote the total transmission time divided into $K+1$ blocks, each of which of duration $\frac{T}{K+1}$ (sec) is allocated to one of the transmitter including the source and $K$ relays, as illustrated in Fig. \ref{fig:PsNcDF}. Each block consists of $N_s$ symbols of duration $T_s$ (sec) each.  In the first block, the source broadcasts its signal with power $P_0$ (watts). {\blue Each relay splits its received RF signal power into two portions: the $\rho$ portion is utilized for WPT and the harvested energy at relay $\mathrm{T}_r$ is $E_r=\eta\rho P_0\mathcal{L}_{0r}|h_{0r}|^2\frac{T}{K+1}$, $r=1,\cdots,K$, where $0<\eta<1$ is the EH efficiency \cite{X.Zhou2013.11}.\footnote{{\blue Note that the inter-relay channel (IRC) is not exploited for EH, because the energy that could be harvested at a relay from any other EH relays via the IRC is negligible as compared to the energy harvested directly from the source.}} The remaining $1-\rho$ portion is for WIT.  Thus, the harvested power $P_r$ (watts) available for data relaying at $\mathrm{T}_r$ is given by 
\begin{align}
  P_r=\frac{E_r}{\frac{T}{K+1}}=\eta \rho P_0\mathcal{L}_{0r}|h_{0r}|^2.\label{eq:Pk_PS}
\end{align}}
In the remaining $K$ blocks, the relays take turns to decode and forward the source information to the destination using full portion of the harvested power $P_r$ in \eqref{eq:Pk_PS}. An extreme case of $\rho = 0$ represents a pure WIT system with no EH, while $\rho = 1$ corresponds to a pure WPT system with no ID. In this paper, we consider a general SWIPT system featuring both WIT and WPT; thus, we have $0<\rho<1$.

\subsubsection{PS-NcDF for $M$-DPSK}
For the $M$-DPSK signaling, the source message $m\in\mathcal{M}\pardef \{0,1,\cdots,M-1\}$ in the $l$-th symbol time, $l=0,1,\cdots$, is conveyed via two consecutive information-bearing symbols $s(l)$ and $s(l-1)$ according to the differential encoding $s(l)=s(l-1)\e^{j2\pi m/M}$, where $s(0)=1$ is the initial reference signal. According to the power splitting, the received RF signal power at each relay is split in the proportion of $\rho$ and $1-\rho$ , where the fractions $\rho$ and $1-\rho$ of the power are utilized for EH and ID, respectively. Thus, the received signal  at $\mathrm{T}_r$, $r=1,\cdots,K$, intended for ID is 
\begin{align}
  y_{0r}(l)=\sqrt{(1-\rho)P_0T_s\mathcal{L}_{0r}}h_{0r}s(l)+\sqrt{1-\rho}u_{0r}(l)+v_{0r}(l),\label{eq:y0r_DPSK}
\end{align}
where $T_s$ (sec) is the symbol time, $u_{0r}(l)\sim\mathcal{CN}(0,\sigma_{0r,1}^2)$ is the AWGN introduced at the receive antenna at $\mathrm{T}_r$, and $v_{0r}(l)\sim\mathcal{CN}(0,\sigma_{0r,2}^2)$ is the AWGN due to the ID circuit at $\mathrm{T}_r$ (which includes the RF-to-baseband conversion noise, the analog-to-digital conversion noise, circuit loss, etc) \cite{K.Huang2013.12,X.Zhou2013.11}. The destination does not have a power splitter and the received signal is solely used for ID. Thus, the received sinal at the destination from the direct source-destination link is given by 
\begin{align}
  y_{0d}(l)=\sqrt{P_0T_s\mathcal{L}_{0d}}h_{0d}s(l)+u_{0d}(l)+v_{0d}(l),\label{eq:y0d_DPSK}
\end{align}
where $u_{0d}(l)\sim\mathcal{CN}(0,\sigma_{0d,1}^2)$ and $v_{0d}(l)\sim\mathcal{CN}(0,\sigma_{0d,2}^2)$ are the AWGNs due to the receive antenna and ID circuit at $\mathrm{T}_d$, respectively. Each relay node assists the communication from the source to the destination via the DF protocol, where $\mathrm{T}_r$ detects the source message by ML detection and forwards the detected message $m_r\in \mathcal{M}$ with full harvested power $P_r$ in \eqref{eq:Pk_PS} to the destination in a time-division manner via the $M$-DPSK transmission. Specifically, the transmitted symbols $s_r(l)$, $l=0,1,\cdots$, at $\mathrm{T}_r$ are successively related according to the differential encoding $s_r(l)=s_r(l-1)\e^{j2\pi m_r/M}$, where $s_r(0)=1$ is the initial reference signal at $\mathrm{T}_r$, $r=1,\cdots,K$. The signal received at the destination from $\mathrm{T}_r$ is given by 
\begin{align}
  y_{rd}(l)=\sqrt{P_rT_s\mathcal{L}_{rd}}h_{rd}s_r(l)+u_{rd}(l)+v_{rd}(l),\label{eq:yrd_DPSK}
\end{align}
where $u_{rd}(l)\sim\mathcal{CN}(0,\sigma_{rd,1}^2)$ and $v_{rd}(l)\sim\mathcal{CN}(0,\sigma_{rd,2}^2)$ are AWGNs due to the receive antenna and ID circuit, respectively. It is assumed that all AWGNs are mutually independent.

\subsubsection{PS-NcDF for $M$-FSK}
For noncoherent $M$-FSK, the source message $m\in \mathcal{M}$ is transmitted over one of the $M$ orthogonal carriers. The received RF signal at each receive terminal is fed to $M$ orthogonal subband correlators, yielding an $M\times 1$ baseband equivalent signal vector $\by_{ij}\pardef[\mathrm{y}_{ij}(1),\cdots,\mathrm{y}_{ij}(M)]^T$, $ij\in\{0d,0r,rd\}_{r=1}^N$. The baseband equivalent signal received at the destination from the direct link is 
\begin{align}
  \by_{0d}=\sqrt{P_0T_s\mathcal{L}_{0d}}h_{0d}\bi_{m+1}+\bu_{0d}+\bv_{0d},\label{eq:by_0d}
\end{align}
where $\bu_{0d}\sim\mathcal{CN}(\bm{0},\sigma_{0d,1}^2\bm{I}_M)$ and $\bv_{0d}\sim\mathcal{CN}(\bm{0},\sigma_{0d,2}^2\bm{I}_M)$ are the AWGNs due to the receive antenna and the ID circuit at $\mathrm{T}_d$, respectively. Due to power splitting at each relay, the received signal intended for ID at $\mathrm{T}_r$ is scaled by the factor $\sqrt{1-\rho}$, and the baseband equivalent received signal at $\mathrm{T}_r$, $r=1,\cdots,K$,  is given by 
\begin{align}
  \by_{0r}=\sqrt{(1-\rho)P_0T_s\mathcal{L}_{0r}}h_{0r}\bi_{m+1}+\sqrt{1-\rho}\bu_{0r}+\bv_{0r},\label{eq:by_0r}
\end{align}
where $\bu_{0r}\sim\mathcal{CN}(\bm{0},\sigma_{0r,1}^2\bm{I}_M)$ and $\bv_{0r}\sim\mathcal{CN}(\bm{0},\sigma_{0r,2}^2\bm{I}_M)$ are AWGNs due to the receive antenna and the ID circuit at $\mathrm{T}_r$, respectively. Each relay decodes the source message and forwards the decoded message $m_r\in\mathcal{M}$  with full power $P_r$ in \eqref{eq:Pk_PS} to the destination in a time-division fashion. The baseband equivalent signal received at the destination from $\mathrm{T}_r$ is given by 
\begin{align}
  \by_{rd}=\sqrt{P_rT_s\mathcal{L}_{rd}}h_{rd}\bi_{m_r+1}+\bu_{rd}+\bv_{rd},\label{eq:by_rd}
\end{align}
where $\bu_{rd}\sim\mathcal{CN}(\bm{0},\sigma_{rd,1}^2\bm{I}_M)$ and $\bv_{rd}\sim\mathcal{CN}(\bm{0},\sigma_{rd,2}^2\bm{I}_M)$ are AWGNs due to receive antenna and ID circuit at $\mathrm{T}_d$, respectively. The AWGNs are assumed mutually independent.

\subsection{TS-NcDF Relaying}\label{sub:TS_NcDF}
{\blue In the TS-NcDF protocol, the WPT and WIT functionalities take place in a time-division fashion. Specifically, the total transmission time $T$ (sec) is divided into two phases: the WPT phase of duration $\alpha T$  and the WIT phase of duration $(1-\alpha) T$, as illustrated in Fig. \ref{fig:TsNcDF}.} {\blue In the WPT phase, the source radiates RF energy flow with power $P_0$ to the relays, where only the EH circuits are activated. Thus, the harvested energy at $\mathrm{T}_r$ is $E_r=\eta P_0\mathcal{L}_{0r}|h_{0r}|^2\alpha T$, $r=1,\cdots,K$, where $0<\eta<1$ is the energy conversion efficiency \cite{X.Zhou2013.11}. In the WIT phase, only the ID circuit is activated at each relay. To coordinate multiple terminals' transmissions, the WIT phase is further divided into $K+1$ blocks of duration $\frac{(1-\alpha)T}{K+1}$ each. In the first block, the source broadcasts information with power $P_0$ to all relays. The relay $\mathrm{T}_r$ utilizes its harvested energy $E_r$ to decode and forward the source information to the destination in the $(r+1)$-th block, $r=1,\cdots,K$. Thus, the harvested power $P_r$ available for data relaying at $\mathrm{T}_r$  is given by
\begin{align}
  P_r=\frac{E_r}{\frac{(1-\alpha)T}{K+1}}=\frac{(K+1)\eta P_0\mathcal{L}_{0r}|h_{0r}|^2\alpha}{1-\alpha}.\label{eq:Pk_TS}
\end{align}}
An extreme case of $\alpha = 0$ represents a pure WIT system with no EH, and $\alpha = 1$ corresponds to a pure WPT system with no ID. In this paper, we consider $0<\alpha<1$, which corresponds to the general SWIPT relay system featuring both WIT and WPT.



\subsubsection{TS-NcDF for $M$-DPSK}

In the TS-NcDF protocol, the signal received at each terminal during the WIT phase is solely utilized for ID, i.e., no power splitting is involved. Consequently, the signal model for TS-NcDF employing the $M$-DPSK is depicted as follows: 
\begin{subequations}\label{eq:yij_DPSK_TS}
  \begin{align}
  y_{0d}(l)&=\sqrt{P_0T_s\mathcal{L}_{0d}}h_{0d}s(l)+u_{0d}(l)+v_{0d}(l),\label{eq:y0d_DPSK_TS}\\
  y_{0r}(l)&=\sqrt{P_0T_s\mathcal{L}_{0r}}h_{0r}s(l)+u_{0r}(l)+v_{0r}(l),\label{eq:y0r_DPSK_TS}\\
  y_{rd}(l)&=\sqrt{P_rT_s\mathcal{L}_{rd}}h_{rd}s_r(l)+u_{rd}(l)+v_{rd}(l),\label{eq:yrd_DPSK_TS}
\end{align}
\end{subequations}
$r=1,\cdots,K$, where $u_{ij}(l)\sim\mathcal{CN}(0,\sigma_{ij,1}^2)$ and $v_{ij}(l)\sim\mathcal{CN}(0,\sigma_{ij,2}^2)$ are AWGNs due to the receive antenna and ID circuit, respectively, $ij\in\{0d,0r,rd\}_{r=1}^K$. The source symbol $s(l)$ is related to the message $m$ through $s(l)=s(l-1)\e^{j2\pi m/M}$. Similarly, $s_r(l)=s_r(l-1)\e^{j2\pi m_r/M}$.

\subsubsection{TS-NcDF for $M$-FSK}
The signal model for TS-NcDF employing noncoherent $M$-FSK is\begin{subequations}\label{eq:by0d_TS}
\begin{align}
  \by_{0d}&=\sqrt{P_0T_s\mathcal{L}_{0d}}h_{0d}\bi_{m+1}+\bu_{0d}+\bv_{0d},\label{eq:by_0d_TS}\\
  \by_{0r}&=\sqrt{P_0T_s\mathcal{L}_{0r}}h_{0r}\bi_{m+1}+\bu_{0r}+\bv_{0r},\label{eq:by_0r_TS}\\
  \by_{rd}&=\sqrt{P_rT_s\mathcal{L}_{rd}}h_{rd}\bi_{m_r+1}+\bu_{rd}+\bv_{rd},\label{eq:by_rd_TS}
\end{align}
\end{subequations}
for $r=1,\cdots,K$, where $\bu_{ij}\sim\mathcal{CN}(\bm{0},\sigma_{ij,1}^2\bm{I}_M)$ and $\bv_{ij}\sim\mathcal{CN}(\bm{0},\sigma_{ij,2}^2\bm{I}_M)$, $ij\in\{0d,0r,rd\}_{r=1}^K$.

\subsection{A Unified Noncoherent SWIPT Framework}\label{sub:unifiedSysModel}
Since we consider two protocols, namely PS-NcDF and TS-NcDF, employing two different noncoherent modulations based on $M$-DPSK or $M$-FSK, there are in total four possible schemes considering different combinations,  which make the derivations of the detectors rather tedious. For ease of development, we now unify the system models of PS-NcDF and TS-NcDF such that the detectors for PS-NcDF and TS-NcDF are derived in a general unified form. The key to the unifying process is a set of reparameterizations of the noise variances and the signal-to-noise ratios (SNRs) as follows:
\begin{subequations}\label{eq:parameters}
\begin{align}
   \gamma_{0d} & \pardef \frac{P_0T_s\mathcal{L}_{0d}}{\sigma_{0d,1}^2 + \sigma_{0d,2}^2}, ~~ \sigma_{kd}^2 \pardef \sigma_{kd,1}^2 + \sigma_{kd,2}^2, ~~
    \sigma_{0r}^2 \pardef \left \{\begin{array}{ll}
                                    (1-\rho)\sigma_{0r,1}^2+\sigma_{0r,2}^2, & \text{PS-NcDF}, \\
                                    \sigma_{0r,1}^2+\sigma_{0r,2}^2, & \text{TS-NcDF},
                                  \end{array}
   \right.\\
   \gamma_{0r} & \pardef \left \{\begin{array}{ll}
                                    \frac{(1-\rho)P_0T_s\mathcal{L}_{0r}}{(1-\rho)\sigma_{0r,1}^2+\sigma_{0r,2}^2}, & \text{PS-NcDF}, \\
                                    \frac{P_0T_s\mathcal{L}_{0r}}{\sigma_{0r,1}^2+\sigma_{0r,2}^2}, & \text{TS-NcDF},
                                  \end{array}
   \right. ~~
   \gamma_{rd}\pardef \left \{\begin{array}{ll}
                                    \frac{\eta\rho P_0T_s\mathcal{L}_{0r}\mathcal{L}_{rd}}{\sigma_{rd,1}^2+\sigma_{rd,2}^2}, & \text{PS-NcDF}, \\
                                    \frac{\alpha (K+1)\eta P_0T_s\mathcal{L}_{0r}\mathcal{L}_{rd}}{(1-\alpha)(\sigma_{rd,1}^2+\sigma_{rd,2}^2)}, & \text{TS-NcDF},
                                  \end{array}
   \right.\label{eq:gamma_rd}
\end{align}
\end{subequations}
where $r = 1,\cdots,K$ and $k = 0,\cdots,K$.
\subsubsection{Unified PS/TS-NcDF Framework for $M$-DPSK}
In the $M$-DPSK transmission, each message is conveyed via two consecutive symbols. Thus, it is convenient to represent two consecutive received
symbols as a vector. To that end, we introduce the following vectors:
\begin{align}
  \bm{y}_{ij}\pardef \left[\begin{array}{c}
                             y_{ij}(l-1) \\
                             y_{ij}(l)
                           \end{array}
  \right], ~~
  \bm{s}\pardef \left[\begin{array}{c}
                             s(l-1) \\
                             s(l)
                           \end{array}
  \right], ~~
  \bm{s}_r\pardef \left[\begin{array}{c}
                             s_r(l-1) \\
                             s_r(l)
                           \end{array}
  \right].\label{eq:DPSK_vector}
\end{align}
Using definitions in \eqref{eq:parameters}, we can unify PS-NcDF and TS-NcDF for $M$-DPSK as follows:
\begin{subequations}\label{eq:DPSK_unified}
\begin{align}
  \bm{y}_{0r}&=\sigma_{0r}\big(\sqrt{\gamma_{0r}}h_{0r}\bm{s}+\bm{n}_{0r}\big),\label{eq:y0r_DPSK_unified}\\
  \bm{y}_{0d}&=\sigma_{0d}\big(\sqrt{\gamma_{0d}}h_{0d}\bm{s}+\bm{n}_{0d}\big),\label{eq:y0d_DPSK_unified}\\
  \bm{y}_{rd}&=\sigma_{rd}\big(\sqrt{\gamma_{rd}}|h_{0r}|h_{rd}\bm{s}_r+\bm{n}_{rd}\big),\label{eq:yrd_DPSK_unified}
\end{align}
\end{subequations}
where $\bm{n}_{ij}\sim\mathcal{CN}(\bm{0},\bm{I}_2)$ for $ij\in\{0d,0r,rd\}_{r=1}^K$. By appropriately choosing the parameters according to \eqref{eq:parameters}, the unified model \eqref{eq:DPSK_unified} can represent either PS-NcDF or TS-NcDF.

\subsubsection{Unified PS/TS-NcDF Framework for Noncoherent $M$-FSK}
Based on the parameters in \eqref{eq:parameters}, the system models of PS-NcDF and TS-NcDF associated with noncoherent $M$-FSK can be expressed in a unified form as follows:
\begin{subequations}\label{eq:FSK_unified}
\begin{align}
  \by_{0r}&=\sigma_{0r}\big(\sqrt{\gamma_{0r}}h_{0r}\bi_{m+1}+\bn_{0r}\big),\label{eq:y0r_FSK_unified}\\
  \by_{0d}&=\sigma_{0d}\big(\sqrt{\gamma_{0d}}h_{0d}\bi_{m+1}+\bn_{0d}\big),\label{eq:y0d_FSK_unified}\\
  \by_{rd}&=\sigma_{rd}\big(\sqrt{\gamma_{rd}}|h_{0r}|h_{rd}\bi_{m_r+1}+\bn_{rd}\big),\label{eq:yrd_FSK_unified}
\end{align}
\end{subequations}
where $\bn_{ij}\sim\mathcal{CN}(\bm{0},\bm{I}_M)$ for $ij\in\{0d,0r,rd\}$ and $r=1,\cdots,K$. {\blue For ease of readability, a list of symbols are summarized in Table \ref{table:symbols}.

Under the noncoherent SWIPT framework, the instantaneous CSIs $h_{ij}$, $ij\in\{0d,0r,rd\}_{r=0}^K$, are not available to the whole network, which makes the noncoherent detection necessary. To ensure the end-to-end optimality of the noncoherent communication, it is required that the destination as well as the relays perform the ML detection. Specifically, each relay performs its own (noncoherent) ML detection individually and forwards its estimated message to the destination. Due to potential decoding errors at the relays, the ultimate (noncoherent)  ML detection at the destination must take into account the transition decoding error probabilities in all the relays, which will be treated in more detail in what follows.

\begin{table}[!t]
\centering \setlength{\tabcolsep}{3.5pt}
{\blue
\caption{List of symbols.}
\begin{tabular}{|c|c||c|c|}
  \hline
   $M$ & Modulation alphabet size & $P_r$, $E_r$ & Harvested power and energy at the $r$-th relay\\ \hline
   $K$ & Number of relays   & $T_s$, $T$  & Symbol time and block time  \\ \hline
   $\eta$  & Energy conversion efficiency & $\mathrm{T}_i$ & The $i$-th terminal \\ \hline
   $\alpha$  & Time switching coefficient & $h_{ij}$, $\mathcal{L}_{ij}$ & Channel coefficient, path loss for the $\mathrm{T}_i$-$\mathrm{T}_j$ link  \\ \hline
   $\rho$  & Power splitting factor & $\sigma_{ij,1}^2$, $\sigma_{ij,2}^2$ & Variances of AWGNs due to receive antenna
and ID circuit  \\ \hline
   $P_0$  & Source transmission power & $\sigma_{ij}^2$, $\gamma_{ij}$ & Noise variance and SNR of $\mathrm{T}_i$-$\mathrm{T}_j$ link   \\ \hline
\end{tabular}\label{table:symbols}}
\end{table}


}

\section{Mathematical Fundamentals}\label{sec:math}
{\blue To facilitate the development of the noncoherent detection schemes, it is necessary to study the distributions involving combinations of CSCG and Rayleigh random variables/vectors, as indicated by the unified SWIPT frameworks in \eqref{eq:DPSK_unified} and \eqref{eq:FSK_unified}.}

\subsection{Generic Probability Analysis}\label{sub:prob_analysis}
We carry out a generic probability analysis involving independent random varaibles/vectors as follows:
\begin{equation}\label{eq:RVs}
\begin{aligned}
  &\bm{x}_0\sim\mathcal{CN}(\bm{0},\Omega_0\bm{I}_2), ~  \bm{y}_0\sim\mathcal{CN}(\bm{0},\Omega_0\bm{I}_M), \\
  &X_i\sim\mathcal{CN}(0,\Omega_i), ~ i=1,2.
\end{aligned}
\end{equation}
We are interested in the following random vectors:
\begin{align}
  \bm{X}_0&\pardef X_1\big|X_2\big|\bm{c} + \bm{x}_0,\label{eq:X0}\\
\bm{Y}_0&\pardef X_1\big|X_2\big|\bi_p+\bm{y}_0,\label{eq:Y0}
\end{align}
where $\bm{c}\pardef [1,c]^T$, $c$ is any complex constant, and $1\leq p \leq M$ is any integer number. Note that \eqref{eq:X0} and \eqref{eq:Y0} are generic forms of $\bm{y}_{rd}$ in \eqref{eq:yrd_DPSK_unified} for $M$-DPSK and $\by_{rd}$ in \eqref{eq:yrd_FSK_unified} for $M$-FSK, respectively.


\begin{lemma}[$\text{CSCG}\times\text{Rayleigh}+\text{CSCG}$]\label{lemma:PDF}
  Given the random variables/vectors in \eqref{eq:RVs}, the PDFs of $\bm{X}_0$ in \eqref{eq:X0} and $\bm{Y}_0$ in \eqref{eq:Y0} are, respectively, given by
\begin{align}
 f_{\bm{X}_0}(\bm{x})&=\frac{\exp\left(-\frac{|x_2-cx_1|^2}{\Omega_0(1+|c|^2)}\right)}{\pi^2\Omega_0^2}I\left(\frac{\Omega_1\Omega_2}
{\Omega_0}(1+|c|^2),\frac{|x_1+c^*x_2|^2}{\Omega_0(1+|c|^2)}\right),\label{eq:dist_X0}\\
f_{\bm{Y}_0}(\bm{y})&=\frac{1}{(\pi\Omega_0)^M}\e^{-\frac{\|\bm{y}\|^2-|y_p|^2}{\Omega_0}}I\left(\frac{\Omega_1\Omega_2}{\Omega_0},\frac{|y_p|^2}{\Omega_0}\right),
\label{eq:dist_Y0}
\end{align}
  where $\bm{x}\pardef [x_1,x_2]^T$, $\bm{y}\pardef [y_1,\cdots,y_M]^T$, and $I(\cdot,\cdot)$ is defined as
  \begin{align}\label{eq:I_def}
    I(\epsilon,\beta)\pardef\frac{\e^{\frac{1}{\epsilon}}}{\epsilon}\int_0^\beta\frac{1}{x}\e^{-\big(x+\frac{\beta}{\epsilon x}\big)}\diff x.
  \end{align}
\end{lemma}
\begin{proof}
  See Appendix A. 
\end{proof}


\begin{corollary}[$\text{CSCG}\times\text{CSCG}+\text{CSCG}$]\label{co:PDF}
  Let $\bm{X}_1\pardef X_1X_2\bm{c}+\bm{x}_0$ and $\bm{Y}_1\pardef X_1X_2\bi_p+\bm{y}_0$. Then the PDF of $\bm{X}_1$, $f_{\bm{X}_1}(\bm{x})$, and the PDF of $\bm{Y}_1$, $f_{\bm{Y}_1}(\bm{y})$, are given by $f_{\bm{X}_1}(\bm{x})=f_{\bm{X}_0}(\bm{x})$ and $f_{\bm{Y}_1}(\bm{y})=f_{\bm{Y}_0}(\bm{y})$.
\end{corollary}
\begin{proof}
  Due to rotational invariance of the CSCG distribution \cite{A.Papoulis2002book}, it can be shown that $X_1X_2$ follows the same distribution as $X_1|X_2|$, which yields the above result.
\end{proof}

A comprehensive study of the distribution of the product of two independent CSCG random variables, namely, ``$\text{CSCG}\times\text{CSCG}$", was also presented in \cite{N.O.D.2012.3}. However, the result in \cite{N.O.D.2012.3} was given in terms of modified bessel function. If one utilizes the result of \cite{N.O.D.2012.3}, then the PDF of ``$\text{CSCG}\times\text{CSCG}+\text{CSCG}$" may be derived as an \emph{infinite} convolution integral of the modified bessel function with the Gaussian PDF, which is prohibitively more complex than the result obtained here. Note that the \emph{finite} integral of \eqref{eq:I_def} has a well-behaved integrand involving only \emph{elementary} functions, and it has a fast convergence speed which allows for fast evaluations at very high accuracy, as will be demonstrated later.


In order to develop the MLD at the destination, the  transition (error) probability of the DPSK detector at the relay needs to be determined. To that end, it is useful to study the distribution of the phase difference of two consecutively received DPSK signals.

\begin{lemma}\label{lemma:phase_dist}
  Given the random variables/vectors in \eqref{eq:RVs}, we consider the $2\times 1$ random vector $\bm{Y}_2\pardef [Y_1,Y_2]^T= X_1\bm{c} + \bm{x}_0$, where $\bm{c}=[1,c]^T$. Then the PDF of $\Theta\pardef\measuredangle Y_1^*Y_2-\measuredangle c$ is
  \begin{align}
    f_{\Theta}(\theta)=\frac{1}{2\pi}\frac{1}{1+\frac{\gamma^2|c|^2\sin^2\theta}{1+\gamma(1+|c|^2)}}\Bigg [1+\frac{\gamma|c|\cos\theta\arccos\Big(-\frac{\gamma|c|\cos\theta}{\sqrt{\gamma^2|c|^2
    +\gamma(1+|c|^2)+1}}\Big)}{\sqrt{\gamma^2|c|^2\sin^2\theta+\gamma(1+|c|^2)+1}}\Bigg],\label{eq:phase_dist}
  \end{align}
where $\gamma\pardef\Omega_1/\Omega_0$ and $\theta$ is defined in \emph{any} $2\pi$ interval of interest.\footnote{Mathematically, $f_{\Theta}(\theta)$ in \eqref{eq:phase_dist} is a periodic function with period $2\pi$. Thus, one may consider the domain of $\theta$ as any $2\pi$ interval of interest such as $[0,2\pi)$ or $[-\pi,\pi)$.}
\end{lemma}
 \begin{proof}
  See Appendix B. 
\end{proof}

The phase difference distribution of PSK/DPSK signals was studied for a simplistic AWGN channel with no fading \cite{R.F.Pawula2001.3}, where the PDF was given in an integral form. In contrast, in addition to the AWGN  $\bm{x}_0\sim\mathcal{CN}(\bm{0},\Omega_0\bm{I}_2)$,  we impose Rayleigh fading where the fading coefficient is modeled by $X_1\sim\mathcal{CN}(0,\Omega_1)$. For the fading scenario, the obtained PDF is given by a truly \emph{closed-form} expression \eqref{eq:phase_dist}, which is more suitable for further analysis.

{\blue The results in Lemma \ref{lemma:PDF} and Corollary \ref{co:PDF} provide a useful characterization of the joint magnitude and phase distribution of sophisticated Gaussian transformations, while the result in Lemma \ref{lemma:phase_dist} characterizes the distribution of the input-output phase difference involving Gaussian random variables/vectors.} These generic analytical results which have not been reported in the literature may be useful in a different context involving different applications.

\subsection{Numerical Computation of $I(\epsilon,\beta)$ in \eqref{eq:I_def}}\label{sub:math}
To facilitate the evaluation of $I(\epsilon,\beta)$ in \eqref{eq:I_def}, we now derive some alternative expressions of $I(\epsilon,\beta)$, which are suitable for numerical computations.
\begin{lemma}[Alternative Forms]\label{lemma:equivalent_forms}
  $I(\epsilon,\beta)$ in \eqref{eq:I_def} can be equivalently expressed as
  \begin{align}
    I(\epsilon,\beta)&=\int_0^\infty\frac{1}{1+\epsilon t}\exp\left(-\frac{\beta}{1+\epsilon t}\right)\exp(-t)\diff t\label{eq:alternative1}\\
    &=\left\{\begin{array}{ll}
        \frac{\e^{\frac{1}{\epsilon}}}{\epsilon}\Bigg[K_0\bigg(2\sqrt{\frac{\beta}{\epsilon}}\bigg)-
        \displaystyle \int_\beta^{\sqrt{\frac{\beta}{\epsilon}}}\frac{1}{x}\e^{-\big(x+\frac{\beta}{\epsilon x}\big)}\diff x \Bigg], & \beta\epsilon<1,\\
        \beta \e^\beta K_0(2\beta), & \beta\epsilon=1, \\
        \frac{\e^{\frac{1}{\epsilon}}}{\epsilon}\Bigg[K_0\bigg(2\sqrt{\frac{\beta}{\epsilon}}\bigg)+
        \displaystyle\int_{\frac{1}{\epsilon}}^{\sqrt{\frac{\beta}{\epsilon}}}\frac{1}{x}\e^{-\big(x+\frac{\beta}{\epsilon x}\big)}\diff x \Bigg], & \beta\epsilon>1,
      \end{array}
    \right.\label{eq:alternative2}
  \end{align}
  where $K_0(x)$ is the zero-order modified Bessel function of the second kind \cite[eq. (8.432.1)]{I.S.Gradshteyn2007book}.
\end{lemma}
\begin{proof}
For $\beta\epsilon<1$, \eqref{eq:I_def} can be rewritten as $I(\epsilon,\beta)=\frac{\e^{\frac{1}{\epsilon}}}{\epsilon}\big [\int_0^{\sqrt{\frac{\beta}{\epsilon}}}\frac{1}{x}\e^{-(x+\frac{\beta}{\epsilon x})}\diff x - \int_\beta^{\sqrt{\frac{\beta}{\epsilon}}}\frac{1}{x}\e^{-(x+\frac{\beta}{\epsilon x})}\diff x \big]$. Similarly, $I(\epsilon,\beta)=\frac{\e^{\frac{1}{\epsilon}}}{\epsilon}\big [\int_0^{\sqrt{\frac{\beta}{\epsilon}}}\frac{1}{x}\e^{-(x+\frac{\beta}{\epsilon x})}\diff x + \int_{\sqrt{\frac{\beta}{\epsilon}}}^\beta \frac{1}{x}\e^{-(x+\frac{\beta}{\epsilon x})}\diff x \big]$ for $\beta\epsilon>1$. For $\beta\epsilon = 1$, we have $I(\epsilon,\beta)=\beta\e^{\beta}\int_0^\beta\frac{1}{x}\e^{-(x+\frac{\beta^2}{x})}\diff x$. These expressions have a common integral-form $\int_0^a\frac{1}{x}\e^{-(x+\frac{a^2}{x})}\diff x$ with $a>0$. By change of variable, it can be shown that $\int_0^a\frac{1}{x}\e^{-(x+\frac{a^2}{x})}\diff x=\int_a^\infty\frac{1}{x}\e^{-(x+\frac{a^2}{x})}\diff x$. Thus, we have $\int_0^a\frac{1}{x}\e^{-(x+\frac{a^2}{x})}\diff x=\frac{1}{2}\int_0^\infty\frac{1}{x}\e^{-(x+\frac{a^2}{x})}\diff x=K_0(2a)$, followed by \cite[eq. (3.471.9)]{I.S.Gradshteyn2007book}. Finally, substituting $a=\sqrt{\beta/\epsilon}$ for $\beta\epsilon>1$ or $\beta\epsilon<1$ and $a=\beta$ for $\beta\epsilon=1$ yields \eqref{eq:alternative2}.
\end{proof}

{\blue The alternative form in \eqref{eq:alternative1} indicates that the original integral of \eqref{eq:I_def} can be seen as the expectation of the random transformation $\frac{1}{1+\epsilon \bm{t}}\exp\left(-\frac{\beta}{1+\epsilon \bm{t}}\right)$, where $\bm{t}$ is exponentially distributed with unit mean. The alternative form in \eqref{eq:alternative2} has a closed-form solution only for $\beta\epsilon=1$. For $\beta\epsilon\neq 1$, we are able to boil down the original integral into two terms, where the first term is given in an exact closed-form while the second term is given by a sub-integral. The benefit of using this alternative form is that the sub-integral is taken over a smaller integral interval and thus can be more easily computed as compared to the original integral in \eqref{eq:I_def}. We now exploit these alternative forms to derive a very useful piecewise approximation for the original integral with very high accuracy.}

\begin{lemma}\label{lemma:numerical_computation}
  For $\beta\epsilon\neq 1$, $I(\epsilon,\beta)$ in \eqref{eq:I_def} can be approximated by the $N$-th order expression,  $I(\epsilon,\beta) = I_N(\epsilon,\beta) + R_N$, $N\geq 1$, where $R_N$ is the $N$-th remainder and $I_N(\epsilon,\beta)$ is given by
\begin{align}
I_N(\epsilon,\beta)\pardef \left\{\begin{array}{ll}
  \tau\e^{-\beta\tau}\bigg[1+\sum\limits_{n=2}^N(-1)^n \epsilon^n\tau^n\mu_n\sum\limits_{\substack{\sum_{i=1}^n im_i=n\\  \forall m_i\in \mathbb{Z}^{+}}}\frac{(\varsigma-\beta\tau)(-\beta\tau)^{\varsigma-1}}{\prod\limits_{j=1}^n m_j!} \bigg], & \beta\epsilon<1, \frac{\e^{\frac{1}{\epsilon}}}{\epsilon}>1,\vspace{2mm}  \\
\frac{\e^{\frac{1}{\epsilon}}}{\epsilon}\bigg[K_0\bigg(2\sqrt{\frac{\beta}{\epsilon}}\bigg)-\Psi_N\bigg(\beta,\sqrt{\frac{\beta}{\epsilon}}\bigg)
        \bigg], & \beta\epsilon<1, \frac{\e^{\frac{1}{\epsilon}}}{\epsilon}\leq 1, \vspace{3mm}  \\
\frac{\e^{\frac{1}{\epsilon}}}{\epsilon}\bigg[K_0\bigg(2\sqrt{\frac{\beta}{\epsilon}}\bigg)+\Psi_N\bigg(\frac{1}{\epsilon},\sqrt{\frac{\beta}{\epsilon}}\bigg)
        \bigg], & \beta\epsilon>1,
\end{array}
\right.\label{eq:N_approx}
\end{align}
 where $\tau\pardef \frac{1}{\epsilon+1}$, $\varsigma\pardef \sum\limits_{i=1}^n m_i$, $\mu_n\pardef \sum\limits_{i=0}^n\frac{(-1)^i}{i!}n!$, and $\mathbb{Z}^{+}\pardef\{0,1,2,\cdots\}$. Also, $\Psi_N(\cdot,\cdot)$ is given by
  \begin{align}
    \Psi_N(a,b)\pardef \frac{b-a}{2}\sum_{i=1}^N w_i\psi\left(\frac{b-a}{2}z_i+\frac{a+b}{2}\right),\label{eq:Psi}
  \end{align}
  where $w_i$ and $z_i$, $i=1,\cdots,N$, are, respectively, the weights and zeros of the Legendre polynomial of order $N$ \cite[Table 25.4]{M.Abramowitz1964book}, and $\psi(x)\pardef\frac{1}{x}\e^{-(x+\frac{\beta}{\epsilon x})}$.
\end{lemma}
\begin{proof}
  See Appendix C. 
\end{proof}

{\blue The first piece of \eqref{eq:N_approx} is a direct consequence of \eqref{eq:alternative1} followed by the Taylor series expansion of $\frac{1}{1+\epsilon \bm{t}}\exp\left(-\frac{\beta}{1+\epsilon \bm{t}}\right)$ around the mean $\mathbb{E}[\bm{t}]=1$. The second and third pieces of \eqref{eq:N_approx} are direct consequences of the Gaussian-Legendre quadrature carried out for the sub-integrals in \eqref{eq:alternative2}.} Note that the closed-form piecewise approximation of $I(\epsilon,\beta)$ by means of $I_N(\epsilon,\beta)$ in \eqref{eq:N_approx} is valid for any natural number $N\geq 1$, and an improved accuracy is expected by increasing the order $N$. However, from the computational cost point-of-view, it is desirable to choose a fairly small order $N$ which attains the required accuracy. To that end, we set $N=2$ to guarantee extremely low computational complexity, and the expression of \eqref{eq:N_approx} reduces to
\begin{subnumcases}{\label{eq:2_approx} I_2(\epsilon,\beta)=}
\tau\e^{-\beta\tau}\Big(1+\epsilon^2\tau^2-2\epsilon^2\tau^3\beta-\frac{1}{2}\epsilon^2\tau^4\beta^2\Big), & $\beta\epsilon<1, \frac{\e^{\frac{1}{\epsilon}}}{\epsilon}>1$,\label{eq:2_approx_a}\\
        \frac{\e^{\frac{1}{\epsilon}}}{\epsilon}\bigg[K_0\bigg(2\sqrt{\frac{\beta}{\epsilon}}\bigg)-\Psi_2\bigg(\beta,\sqrt{\frac{\beta}{\epsilon}}\bigg)
        \bigg], & $\beta\epsilon<1, \frac{\e^{\frac{1}{\epsilon}}}{\epsilon}\leq 1$, \label{eq:2_approx_b}\\
        \frac{\e^{\frac{1}{\epsilon}}}{\epsilon}\bigg[K_0\bigg(2\sqrt{\frac{\beta}{\epsilon}}\bigg)+\Psi_2\bigg(\frac{1}{\epsilon},\sqrt{\frac{\beta}{\epsilon}}\bigg)
        \bigg], & $\beta\epsilon>1$,\label{eq:2_approx_c}
\end{subnumcases}
where we have $z_1=\frac{\sqrt{3}}{3}$, $z_2=-\frac{\sqrt{3}}{3}$, and $w_1=w_2=1$ for $N=2$. Through extensive evaluations in Section \ref{sec:numerical}, we will demonstrate that the piecewise approximation $I_2(\epsilon,\beta)$ in \eqref{eq:2_approx} is indeed very accurate for a wide range of $\epsilon$ and $\beta$ values. Considering its very low computational cost, we will utilize $I_2(\epsilon,\beta)$ in the next section to facilitate the implementation of the approximate MLDs.

\section{MLDs for Energy Harvesting DF Networks}\label{sec:MLD}
Following the unified noncoherent SWIPT framework in Section \ref{sec:NcDF_relaying}, we will develop unified  noncoherent detectors in this section. Specifically, we begin by finding the exact MLDs at each relay, followed by the transition probability analysis. Finally, the exact MLDs as well as the closed-form approximate MLDs at the destination are derived for both $M$-DPSK and $M$-FSK.

\subsection{Unified MLDs at the Relays}
From the unified first-hop model in \eqref{eq:y0r_DPSK_unified} and \eqref{eq:y0r_FSK_unified}, we see that the input-output relation is equivalent to the conventional P2P communication in Rayleigh fading. Thus, the MLD at the relay can be derived similar to that for P2P communication. Specifically, the MLDs of $M$-DPSK at $\mathrm{T}_r$, $r=1,\cdots,K$, for PS-NcDF and TS-NcDF are expressed in a unified form \cite[Ch. 5.2.8]{J.G.Proakis2000book}
\begin{align}
  m_r=\arg\max_{m=0,\cdots,M-1}\Re\big\{y_{0r}^*(l)y_{0r}(l-1)\e^{j2\pi m/M}\big\}.\label{eq:MLD_relay_DPSK}
\end{align}
Similarly, the MLDs of $M$-FSK  at $\mathrm{T}_r$  for PS-NcDF and TS-NcDF are given by \cite[Ch. 5.4.2]{J.G.Proakis2000book}
\begin{align}
  m_r=\arg\max_{m=0,\cdots,M-1}\big|\mathrm{y}_{0r}(m+1)\big|^2.\label{eq:MLD_relay_FSK}
\end{align}

\subsection{Transition Probability Analysis}
Due to potential detection errors at the relay, the detected message $m_r$ at $\mathrm{T}_r$ may be different from the actual source transmitted message $m$. In fact, a potential decision error at $\mathrm{T}_r$ may transform the correct message $m\in \mathcal{M}$ to any incorrect message $m_r\in \mathcal{M}$. Thus, the \emph{transition probability}, $\Pr(m_r|m)$, namely the probability that the message $m_r$ is detected while $m$ is actually transmitted serves as useful side information for the destination detection. With orthogonal signalings such as $M$-FSK, the transition from the correct message $m$ to any one of the $M-1$ incorrect message is equiprobable \cite{J.G.Proakis2000book}. Thus, the transition probability for $M$-FSK can be expressed in terms of the SER, $\varepsilon_r^{\rm f}$, as follows:
\begin{align}
  \Pr(m_r|m) = \left\{\begin{array}{cc}
                        1-\varepsilon_r^{\rm f}, & m_r=m, \\
                        \frac{\varepsilon_r^{\rm f}}{M-1}, & m_r\neq m,
                      \end{array}
  \right.\label{eq:TP_FSK}
\end{align}
where
\begin{align}
  \varepsilon_r^{\rm f}\pardef \sum_{m=1}^{M-1}(-1)^{m+1}{M-1 \choose m}\frac{1}{1+m(1+\gamma_{0r})}.\label{eq:FSK_relay_SER}
\end{align}
For non-orthogonal signalings such as $M$-DPSK, however, the transition errors are not equiprobable because each constellation point is not equidistant from  all other constellation points \cite{J.G.Proakis2000book}. Thus, the simple relationship in \eqref{eq:TP_FSK} for $M$-FSK does not hold for $M$-DPSK. With non-trivial derivations, we derive the \emph{exact} transition probability for $M$-DPSK as follows.

\begin{lemma}[Exact Transition Probability for $M$-DPSK]\label{lemma:transition_prob}
  For the MLD of $M$-DPSK in \eqref{eq:MLD_relay_DPSK} at relay $\mathrm{T}_r$, $r=1,\cdots,K$, the exact transition probability in Rayleigh fading is given by $\Pr(m_r|m)=\mathcal{P}_{|m_r-m|}(\gamma_{0r})$, where
  \begin{align}
  \!\!\!\!  \mathcal{P}_n(\gamma_{0r})=\frac{1}{2\pi}\int_{\frac{2n-1}{M}\pi}^{\frac{2n+1}{M}\pi}\frac{1}{1+\frac{\gamma_{0r}^2\sin^2\theta}{1+2\gamma_{0r}}}\bigg [1+\frac{\gamma_{0r} \cos\theta}{\sqrt{\gamma_{0r}^2\sin^2\theta+2\gamma_{0r}+1}}\arccos\bigg(-\frac{\gamma_{0r}\cos\theta}{1+\gamma_{0r}}\bigg)\bigg]\diff \theta,\label{eq:transition_prob}
  \end{align}
$n=0,1,\cdots,M-1$.
\end{lemma}
\begin{proof}
See Appendix D. 
\end{proof}

{\blue The exact transition probability of $M$-DPSK depends only on the transition phase indexed by $|m_r-m|$.} It is not hard to mathematically prove $\mathcal{P}_n(\gamma_{0r})=\mathcal{P}_{M-n}(\gamma_{0r})$ for $n=0,\cdots,M-1$. This property can be attributed to the symmetry of the PSK constellations. It suggests that in order to characterize $M$ transition probabilities for $M\geq 4$, one only needs to calculate $M/2$ transition probabilities $\mathcal{P}_n(\gamma_{0r})$ for $n=0,\cdots,\frac{M}{2}-1$ via \eqref{eq:transition_prob}, and the remaining $M/2$ transition probabilities, $\mathcal{P}_n(\gamma_{0r})$, $n=\frac{M}{2},\cdots,M$, follow immediately by the symmetry $\mathcal{P}_n(\gamma_{0r})=\mathcal{P}_{M-n}(\gamma_{0r})$.

\begin{lemma}[Closed-Form Approx. Transition Probability]\label{lemma:transition_prob_approx}
  The transition probability of $M$-DPSK in Rayleigh fading can be approximated as
  \begin{equation}\label{eq:TP_approx_by_SER}
  \begin{aligned}
    \Pr(m_r|m)&\approx\left\{\begin{array}{ll}
                               1-\varepsilon^{\text{p}}, & m_r=m,\\
                               \frac{\mathrm{\varepsilon^{\text{p}}}}{2}, & |m_r-m|=1 ~ \text{or} ~ M-1,\\
                               0, & \text{otherwise},
                             \end{array}
    \right.
  \end{aligned}
  \end{equation}
where
\begin{align}\label{eq:PSK_relay_SER}
  \varepsilon_r^{\rm p}\pardef  \left\{\begin{array}{ll}
    \frac{1}{2(1+\gamma_{0r})}, & M=2, \\
     1.03\sqrt{\frac{1+\cos\frac{\pi}{M}}{2\cos\frac{\pi}{M}}}\left[1-\sqrt{\frac{\big(1-\cos\frac{\pi}{M}\big)\gamma_{0r}}{1+
     \big(1-\cos\frac{\pi}{M}\big)\gamma_{0r}}}\right], & M\geq 4.
                                 \end{array}
  \right.
\end{align}
\end{lemma}
\begin{proof}
See Appendix E.
\end{proof}

{\blue The approximate transition probability in \eqref{eq:TP_approx_by_SER} exploits the fact that the error transmissions from the transmitted PSK phase to its two closest neighbouring phases are most likely and euqi-probable. Moreover, error transitions to any farther constellations points other than the two closest constellation points are negligible. }

\subsection{Unified MLDs at the Destination}
In this subsection, we develop the MLDs at the destination for both PS-NcDF and TS-NcDF protocols in a unified form, which exploits explicitly the transition probabilities at the relays to ameliorate the error propagations from the relays.

\begin{theorem}[Exact MLDs]\label{thm:Exact_MLD}
The exact MLDs of $M$-DPSK at the destination for both PS-NcDF and TS-NcDF are given in a unified form as follows:
\begin{align}
 \!\!\!\!\!\! \hat{m} = \arg\!\!\!\!\!\!\max_{m=0,\cdots,M-1}\!\!\Bigg\{\frac{2\gamma_{0d}}{1+2\gamma_{0d}}\beta_0(m)+\sum_{r=1}^K\ln\!\!\Bigg[\sum_{m_r=0}^{M-1}\mathcal{P}_{|m_r-m|}
  (\gamma_{0r})\e^{-\beta_r^{-}(m_r)}I\left(2\gamma_{rd},\beta_r^{+}(m_r)\right)\!\!\Bigg]\!\Bigg\},\label{eq:MLD_DPSK}
\end{align}
where $\beta_0(m)\pardef \frac{\Re\{y_{0d}(l-1)y_{0d}^*(l)\e^{j2\pi m/M}\}}{\sigma_{0d}^2}$, $\beta_r^{+}(m_r)\pardef \frac{|y_{rd}(l)+y_{rd}(l-1)\e^{j2\pi m_r/M}|^2}{2\sigma_{rd}^2}$, $\beta_r^{-}(m_r)\pardef \frac{|y_{rd}(l)-y_{rd}(l-1)\e^{j2\pi m_r/M}|^2}{2\sigma_{rd}^2}$, and $\mathcal{P}_{|m_r-m|}(\gamma_{0r})$ is the transition probability given in \eqref{eq:transition_prob}.
Similarly, the exact MLDs of $M$-FSK at the destination for PS-NcDF and TS-NcDF are given in a unified form as
\begin{align}
  &\hat{m} = \arg\max_{m=0,\cdots,M-1}\Bigg\{\frac{\gamma_{0d}}{1+\gamma_{0d}}\frac{|\mathrm{y}_{0d}(m+1)|^2}{\sigma_{0d}^2}+\sum_{r=1}^K\ln\Bigg[(1-
  \varepsilon_r^{\rm f})\e^{-\frac{\|\by_{rd}\|^2-|\mathrm{y}_{rd}(m+1)|^2}{\sigma_{rd}^2}}\nonumber \\
 \times & I\left(\gamma_{rd},\frac{|\mathrm{y}_{rd}(m+1)|^2}{\sigma_{rd}^2}\right)
    + \frac{\varepsilon_r^{\rm f}}{M-1}\sum_{\substack{m_r=0\\ m_r\neq m}}^{M-1}\e^{-\frac{\|\by_{rd}\|^2-|\mathrm{y}_{rd}(m_r+1)|^2}{\sigma_{rd}^2}}I\left(\gamma_{rd},\frac{|\mathrm{y}_{rd}(m_r+1)|^2}{\sigma_{rd}^2}\right)\Bigg]\Bigg\}.
   \label{eq:MLD_FSK}
\end{align}
\end{theorem}\begin{proof}
 See Appendix F. 
\end{proof}

The exact (optimum) MLDs in \eqref{eq:MLD_DPSK} and \eqref{eq:MLD_FSK} require the evaluations of the integral $I(\epsilon,\beta)$ in \eqref{eq:I_def}. In addition, the transition probability involved in the exact MLD of \eqref{eq:MLD_DPSK} is given by a finite integral-form in \eqref{eq:transition_prob}, which makes the exact MLDs complex to implement in practice. Nevertheless, it is very meaningful to study the exact MLDs as they serve as the optimum performance benchmark for SWIPT in noncoherent DF systems. With the exact MLDs, one may use them as a theoretical performance upper bound to evaluate the performance of suboptimum detectors, if developed. In the following, to make the MLDs suitable for practical systems, we develop \emph{closed-form} approximations for the MLDs.

\begin{theorem}[Approximate MLDs]\label{th:Approx_MLD}
  For $M$-DPSK, the unified-form approximate MLDs for both PS-NcDF and TS-NcDF are given by
\begin{align}\label{eq:approx_MLD_DPSK}
\!\!\!  \hat{m} \approx \left\{\begin{array}{l}
    \arg\max\limits_{m=0,\cdots,M-1}\bigg\{\frac{2\gamma_{0d}}{1+2\gamma_{0d}}\beta_0(m)+\sum\limits_{r=1}^K\ln\Big[(1-\varepsilon_r^{\rm p})\e^{-\beta_r^{-}(m)}
  I_2\big(2\gamma_{rd},\beta_r^{+}(m)\big)\\
  +\varepsilon_r^{\rm p}\e^{-\beta_r^{-}(m+1)}I_2\big(2\gamma_{rd},\beta_r^{+}(m+1)\big)\Big]\bigg\}, ~~~~~~~~~~~~~~~~~~~~~~~~~~~~~~~~~~~~~~~~ M=2,\\
  \arg\max\limits_{m=0,\cdots,M-1}\bigg\{\frac{2\gamma_{0d}}{1+2\gamma_{0d}}\beta_0(m)+\sum\limits_{r=1}^K\ln\Big[(1-\varepsilon_r^{\rm p})\e^{-\beta_r^{-}(m)}
  I_2\big(2\gamma_{rd},\beta_r^{+}(m)\big)\\
  +\varepsilon_r^{\rm p}\frac{\e^{-\beta_r^{-}(m+1)}I_2\big(2\gamma_{rd},\beta_r^{+}(m+1)\big)+
  \e^{-\beta_r^{-}(m-1)}I_2\big(2\gamma_{rd},\beta_r^{+}(m-1)\big)}{2}\Big]\bigg\}, ~~~~~~~~~~~~~~~~~~ M\geq 4. \\
                         \end{array}
  \right.
\end{align}
For $M$-FSK, the approximate MLDs for both PS-NcDF and TS-NcDF are unified as
\begin{align}
  & \hat{m} \approx \arg\max_{m=0,\cdots,M-1} \Bigg\{\frac{\gamma_{0d}}{1+\gamma_{0d}}\frac{|\mathrm{y}_{0d}(m+1)|^2}{\sigma_{0d}^2}+\sum_{r=1}^K\ln\Bigg[(1-
  \varepsilon_r^{\rm f})\e^{-\frac{\|\by_{rd}\|^2-|\mathrm{y}_{rd}(m+1)|^2}{\sigma_{rd}^2}}\nonumber \\
\!\!\!\!\!\! \times &  I_2\left(\gamma_{rd},\frac{|\mathrm{y}_{rd}(m+1)|^2}{\sigma_{rd}^2}\right)
    + \frac{\varepsilon_r^{\rm f}}{M-1}\sum_{\substack{m_r=0\\ m_r\neq m}}^{M-1}\e^{-\frac{\|\by_{rd}\|^2-|\mathrm{y}_{rd}(m_r+1)|^2}{\sigma_{rd}^2}}I_2\left(\gamma_{rd},\frac{|\mathrm{y}_{rd}(m_r+1)|^2}{\sigma_{rd}^2}
   \right)\Bigg]\Bigg\}.
   \label{eq:approx_MLD_FSK}
\end{align}
\end{theorem}
\begin{proof}
Following the exact MLDs in Theorem \ref{thm:Exact_MLD}, applying the approximation of $I(\epsilon,\beta)\approx I_2(\epsilon,\beta)$ for $M$-FSK and $M$-DPSK and the transition probability approximation in Lemma \ref{lemma:transition_prob_approx} for $M$-DPSK  immediately gives the approximate MLDs.
\end{proof}

{\blue The ML decision metrics for the exact MLDs in Theorem \ref{thm:Exact_MLD} are given by the weighted sum of the decision statistics associated with the direct link and the relay branches. For each relay branch, the weight function is proportional to the transition probability and the decision statistic is involved in the integral function $I(\epsilon,\beta)$. For the approximate MLDs in Theorem \ref{th:Approx_MLD}, the exact integral function $I(\epsilon,\beta)$ is replaced by its tight closed-form approximation $I_2(\epsilon,\beta)$. In addition, for $M$-DPSK, the approximate transition probability is used in place of the exact transition probability, which reduces the sum of $M-1$ transition terms to 2 terms, irrespective of $M$. }

\section{Numerical Results}\label{sec:numerical}
In this section, we carry out Monte Carlo simulations to gain useful design insights into the proposed PS-NcDF and TS-NcDF protocols. The EH efficiency at the  relays is set to $\eta = 0.6$ as in \cite{X.Zhou2013.11}. We adopt the \emph{bounded path-loss model} as in \cite{Z.Ding2014.8} in which the path-loss is strictly larger than one. Specifically, the path-loss from transmit terminal $i$ to receive terminal $j$ is modeled by $\mathcal{L}_{ij}=\frac{1}{1+D_{ij}^\varrho}$, where $\varrho$ is the path-loss exponent and $D_{ij}$ is the transmitter-receiver distance for $ij\in\{0d,0r,rd\}_{r=1}^K$. The path-loss exponent is set to $\varrho=2.7$ which corresponds to the urban cellular communication environment \cite{T.S.Rappaport2002book}. The source-destination distance is fixed to $D_{0d}=3$ (m), while the source-relay distance $D_{0r}$ and relay-destination distance  $D_{rd}$ can vary arbitrarily between 0 and 3 subject to $D_{0r}+D_{rd}=3$, for $r=1,\cdots,K$. For ease of simulations, the noise variances are set to be $\sigma_{ij,1}^2=\sigma_{ij,2}^2\pardef\sigma_0^2/2$, such that each receive terminal has a total noise variance $\sigma_0^2$, i.e., $\sigma_{ij,1}^2+\sigma_{ij,2}^2=\sigma_0^2$, where $\sigma_{ij,1}^2$ and $\sigma_{ij,2}^2$, $ij\in\{0d,0r,rd\}_{r=1}^K$, are the variances of the receive antenna noise and the ID circuit noise, respectively. The error performance is parameterized by the \emph{transmitter SNR}, defined as $\mathrm{SNR}\pardef P_0/\sigma_0^2$.

{\blue We will evaluate the error performance of PS-NcDF and TS-NcDF at the same information rate $R\pardef \frac{N_s\log_2 M}{T}$ (bps), where $N_s$ is the total number of information-bearing symbols transmitted over $T$ (sec). Note that $N_s$ new symbols are transmitted in the first block, which are relayed in the following $K$ blocks by the relays. The symbol time $T_s$ is related to $R$  as
\begin{align}
  T_s=\left\{
        \begin{array}{ll}
          \frac{\log_2M}{(K+1)R}, & \hbox{PS-NcDF}, \\
          \frac{(1-\alpha)\log_2M}{(K+1)R}, & \hbox{TS-NcDF}.
        \end{array}
      \right.\label{eq:Ts_Ns}
\end{align}
For simplicity, we will fix $N_s/T=1$  such that $R=\log_2M$, which depends only on $M$.
}

\subsection{Validation of the Piecewise Approximation $I_2(\epsilon,\beta)$ in \eqref{eq:2_approx}}

\begin{figure}[t]%
\centering
\includegraphics[width=0.4\columnwidth]{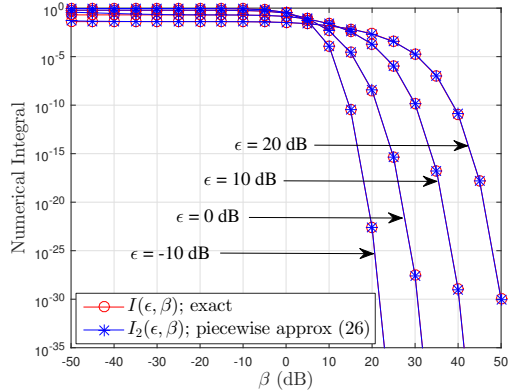}
\caption{Piecewise closed-form approximation $I_2(\epsilon,\beta)$ vs the exact numerical integration $I(\epsilon,\beta)$ for various $\epsilon$ and $\beta$ values.}
\label{fig:Integral_Approx_vs_beta}
\end{figure}


We first check  the accuracy of the proposed piecewise approximation $I_2(\epsilon,\beta)$ in \eqref{eq:2_approx}. As a benchmark, we use the MATLAB numerical integration function ``integral" to directly compute $I(\epsilon,\beta)$ in \eqref{eq:I_def} at a relative error tolerance of $10^{-12}$, which is referred to as the ``exact" result. We evaluate $I_2(\epsilon,\beta)$  for $\beta \in [-50, 50]$ dB under various $\epsilon$ values of interest ranging from $-10$ to $20$ dB in Fig. \ref{fig:Integral_Approx_vs_beta}. Since $\beta$ spans over a very wide range of values for each $\epsilon$, the three subcases involved in \eqref{eq:2_approx}  are all considered in Fig. \ref{fig:Integral_Approx_vs_beta} for each $\epsilon$ value. In this case, the piecewise approximation of \eqref{eq:2_approx} is applied according to the conditions for the subcases. It can be observed that $I_2(\epsilon,\beta)$ and $I(\epsilon,\beta)$ overlap for all possible values of $\epsilon$ and $\beta$, which confirms that $I_2(\epsilon,\beta)$ in \eqref{eq:2_approx} is indeed a very good approximation.

\subsection{Validation of the Approximate MLDs in \eqref{eq:approx_MLD_DPSK} and \eqref{eq:approx_MLD_FSK}}
\begin{figure}[!t]%
\centering
\subfigure[][]{\includegraphics[width=0.4\columnwidth]{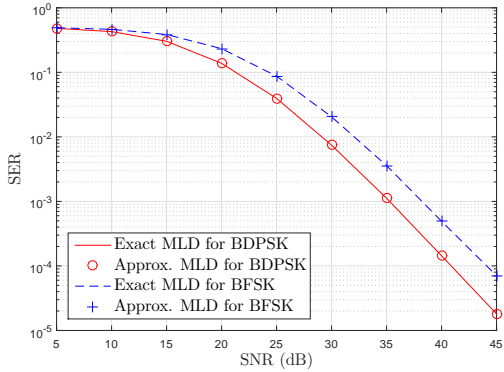}\label{fig:MLDs_TS_Single_Relay}} \hfil
\subfigure[][]{\includegraphics[width=0.4\columnwidth]{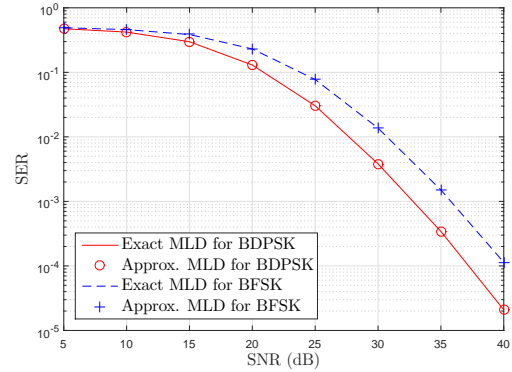}\label{fig:MLDs_PS_Two_Relay}}
\caption{SERs of MLDs for binary signalings ($M=2$) with {\blue $R=1$ (bps)}. \subref{fig:MLDs_TS_Single_Relay} TS-NcDF with $\alpha=0.5$ in a single-relay network ($K=1$) with $D_{0r} = 1.5$ (m). \subref{fig:MLDs_PS_Two_Relay} PS-NcDF with $\rho=0.5$ in a two-relay network ($K=2$) with $D_{01} = 1$ (m) and $D_{02} = 2$ (m)}.
\label{fig:MLDs_comparison_benchamrk}
\end{figure}

We validate the proposed closed-form approximate MLDs in \eqref{eq:approx_MLD_DPSK} and \eqref{eq:approx_MLD_FSK} through comparisons to the exact (integral-form) MLDs in \eqref{eq:MLD_DPSK} and \eqref{eq:MLD_FSK}, for binary DPSK (BDPSK) and binary FSK (BFSK) signalings with fixed {\blue symbol rate $R=1$ (bps)}. Specifically, the TS-NcDF protocol with time switching coefficient $\alpha=0.5$ in a single-relay ($K=1$) EH system with $D_{0r}=1.5$ (m) is considered in  Fig. \ref{fig:MLDs_TS_Single_Relay}, and the PS-NcDF protocol with power splitting factor $\rho=0.5$ for a two-relay ($K=2$) EH system with $D_{01}=1$ (m) and $D_{02}=2$ (m)  is considered  in  Fig. \ref{fig:MLDs_PS_Two_Relay}. As demonstrated in these figures, the SERs of the approximate MLDs are in excellent agreement with those of the exact MLDs over the whole SNR range, indicating that the proposed closed-form approximate MLDs are useful practical solutions for noncoherent SWIPT.

\begin{figure}[!t]%
\centering
\subfigure[][]
{\includegraphics[width=0.4\columnwidth]{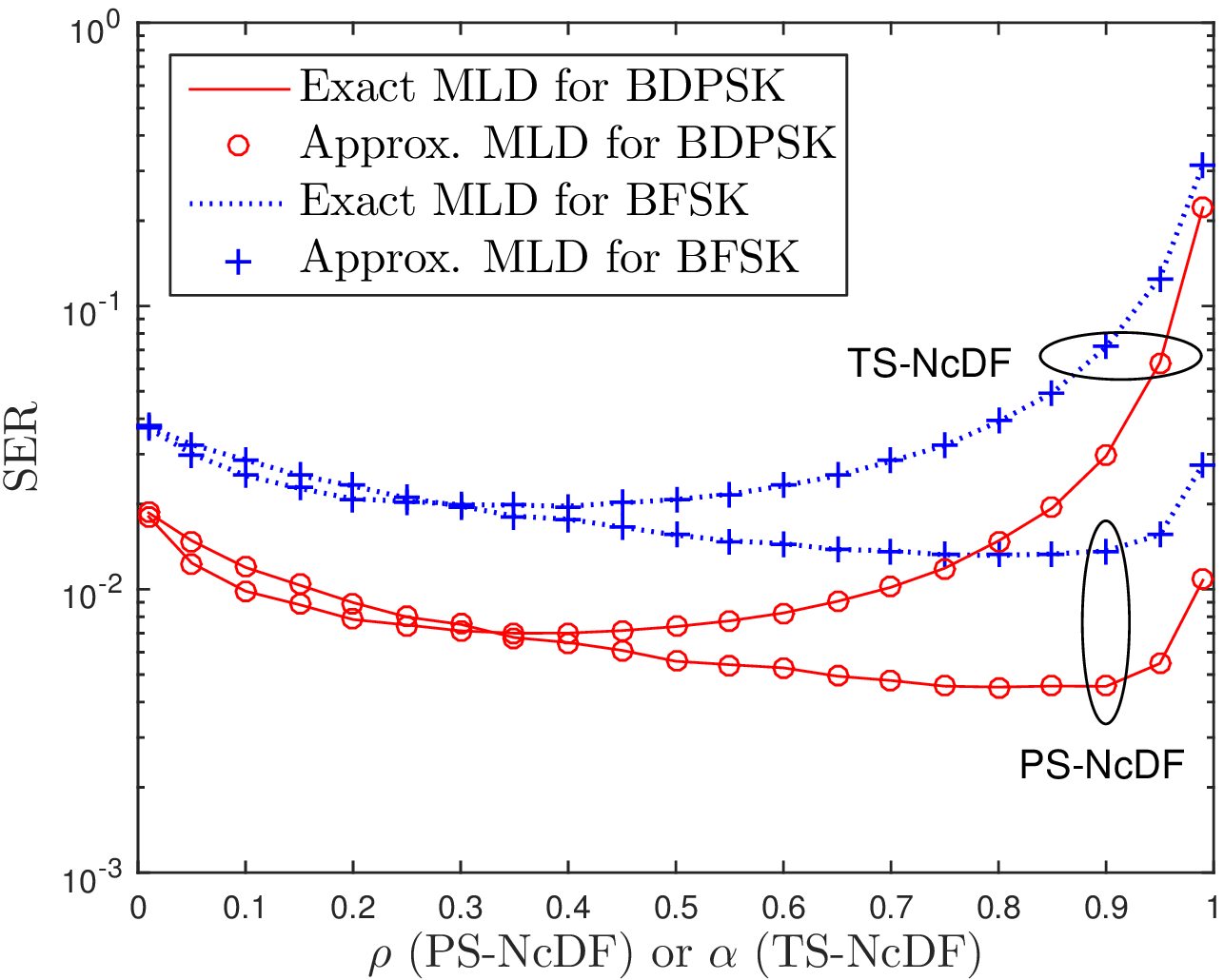}\label{fig:PS_TS_vary_coeff_30dB_M2_fixed_Ns}}
\hfil
\subfigure[][]
{\includegraphics[width=0.4\columnwidth]{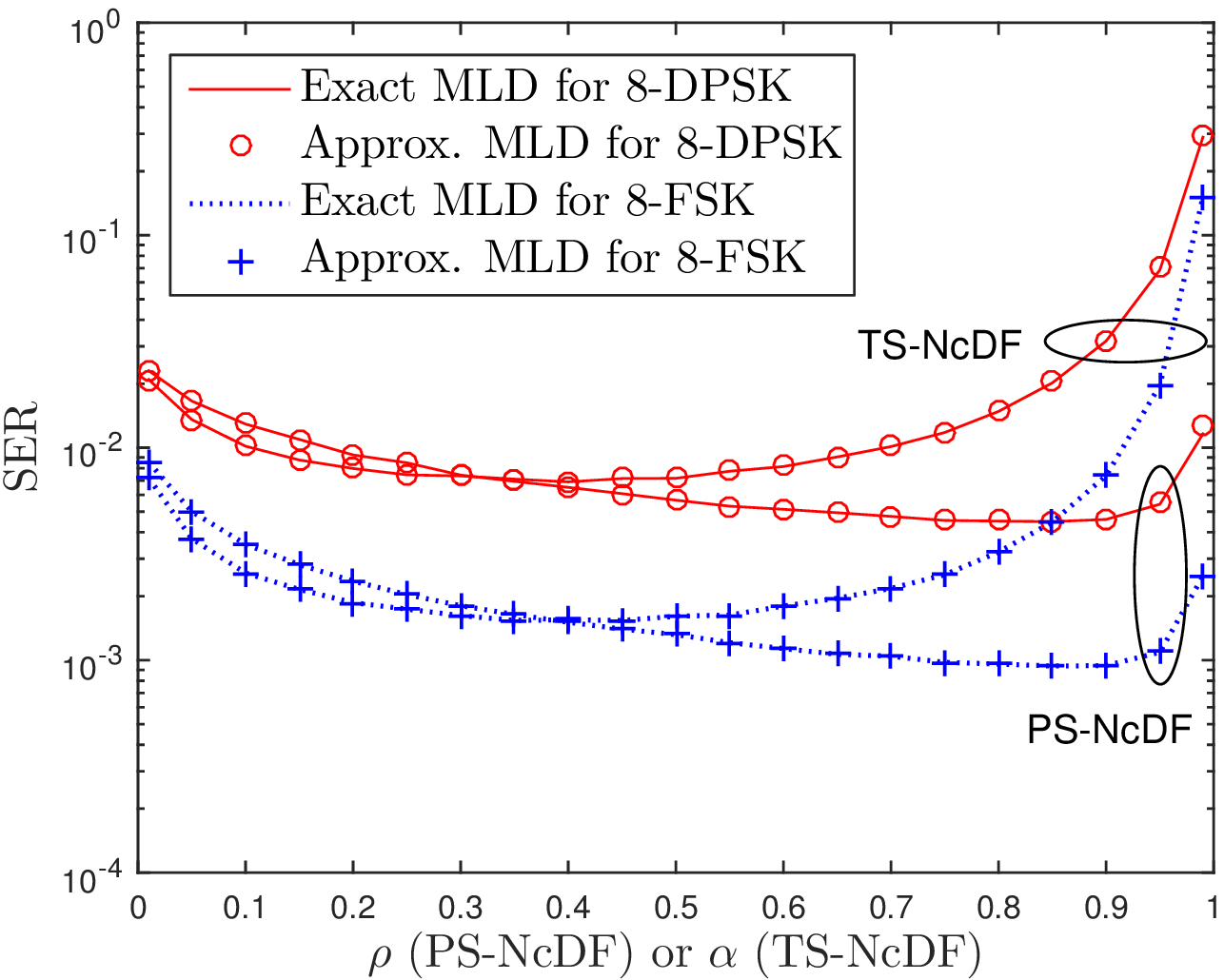}\label{fig:PS_TS_vary_coeff_40dB_M8_fixed_Ns_SNR}}
\caption{Comparison of PS-NcDF  and TS-NcDF for fixed information rate in a single-relay network ($K=1$) with $D_{0r}=1.5$ (m):
\subref{fig:PS_TS_vary_coeff_30dB_M2_fixed_Ns} $M=2$, $\mathrm{SNR} = 30$ dB, and $R = 1$ (bps); and
\subref{fig:PS_TS_vary_coeff_40dB_M8_fixed_Ns_SNR} $M=8$, $\mathrm{SNR} = 40$ dB, and $R = 3$ (bps).
}
\label{fig:PS_TS_fixed_Ns}
\end{figure}

\subsection{Comparison of PS-NcDF and TS-NcDF}
We now compare PS-NcDF and TS-NcDF in terms of the SER {\blue at the same information rate $R$}.  Fig. \ref{fig:PS_TS_fixed_Ns} shows the SER versus the time switching coefficient $\alpha$ for TS-NcDF or power splitting factor $\rho$ for PS-NcDF in a single-relay ($K=1$) network with $D_{0r}=1.5$ (m), where binary signalings ($M=2$) with $R=1$ (bps) and $\mathrm{SNR}=30$ dB (see Fig. \ref{fig:PS_TS_vary_coeff_30dB_M2_fixed_Ns}) and octonary signalings ($M=8$) with $R=3$ (bps) and $\mathrm{SNR}=40$ dB (see Fig. \ref{fig:PS_TS_vary_coeff_40dB_M8_fixed_Ns_SNR}) are considered. As shown in Fig. \ref{fig:PS_TS_fixed_Ns}, the SERs exhibit some global minima around $\alpha=0.4$ for TS-NcDF and around $\rho=0.8$ for PS-NcDF. This is because there exists a performance tradeoff between the first and second hops through the adjustment of $\alpha$ (for TS-NcDF) or $\rho$ (for PS-NcDF). Specifically, when $\alpha$ (or $\rho$) increases, more power (proportional to $\alpha$ or $\rho$) can be harvested and utilized by the second-hop data relaying, resulting in an increased SNR over the second-hop. On the other hand, for TS-NcDF, as $\alpha$ increases, less time (proportional to $1-\alpha$) is available for the first-hop information transmission. Thus, the symbol time $T_s$ for TS-NcDF must decrease so as to maintain the same target information rate $R$, which results in smaller received symbol energy and consequently decreased SNR over the first-hop. Similarly, for PS-NcDF, as $\rho$ increases, less power (proportional to $1-\rho$) is split into the WIT circuit for information decoding; thus, the first-hop SNR decreases. Therefore, adjusting the values of $\alpha$ (or $\rho$) results in different impacts on the SNRs over the two hops, and consequently, a tradeoff exists between the first- and second-hop SNRs through the adjustment of $\alpha$ (or $\rho$). This can be also explained mathematically. By substituting \eqref{eq:Ts_Ns} into \eqref{eq:gamma_rd}, we see that $\gamma_{0r}$ is monotonically decreasing in $\alpha$ (or $\rho$) and $\gamma_{rd}$ is monotonically increasing in $\alpha$ (or $\rho$). Due to the opposite behaviors of the first- and second-hop SNRs as a function of $\alpha$ (or $\rho$), the optimal value of $\alpha$ (or $\rho$) minimizing the overall error probability is strictly between 0 and 1.

\begin{figure}[!t]%
\centering
\subfigure[][$M=2$]
{\includegraphics[width=0.4\columnwidth]{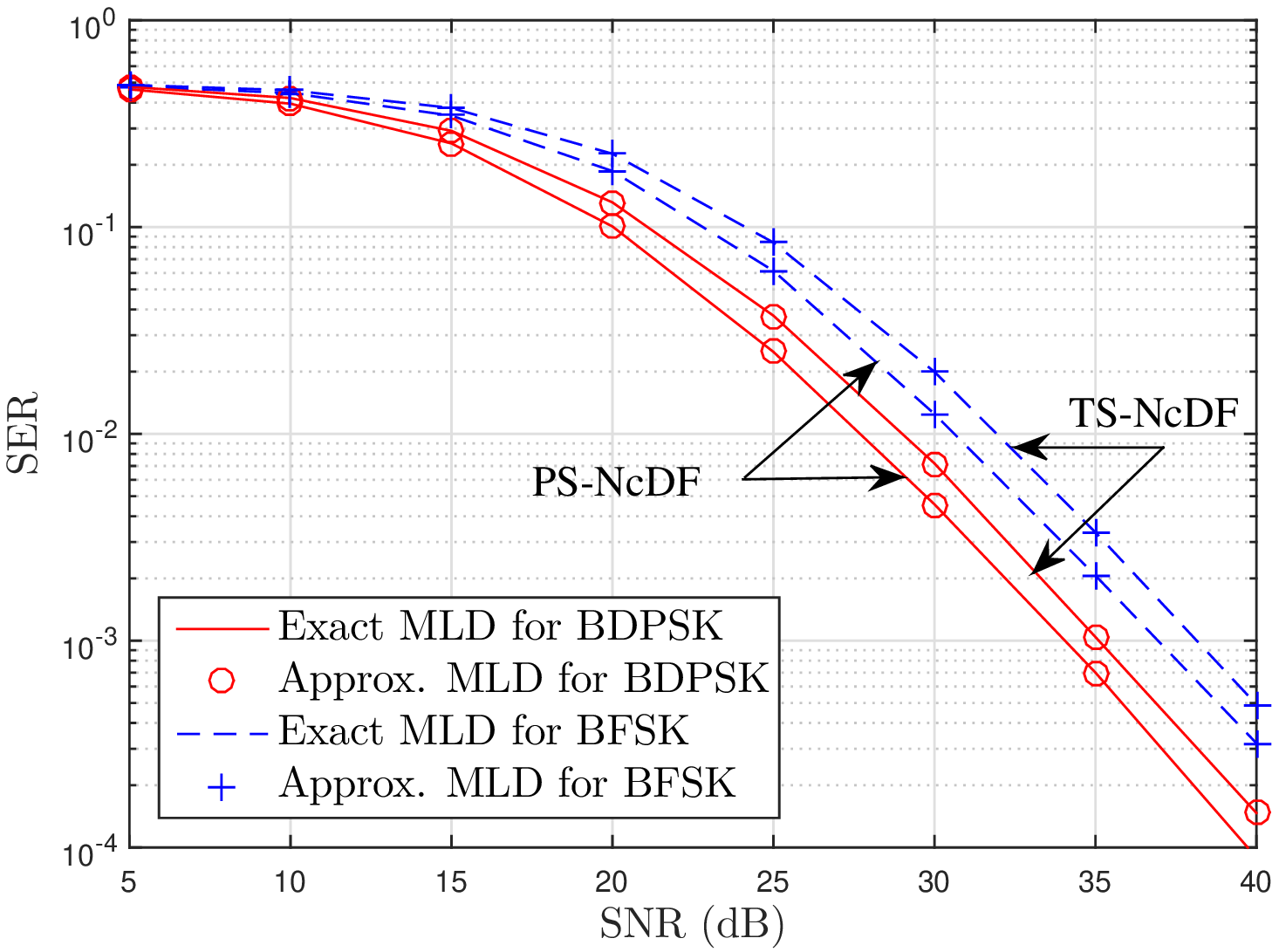}\label{fig:PS_vs_TS_M2_fixed_Ns_SNR}}
\hfil
\subfigure[][$M=4$]
{\includegraphics[width=0.4\columnwidth]{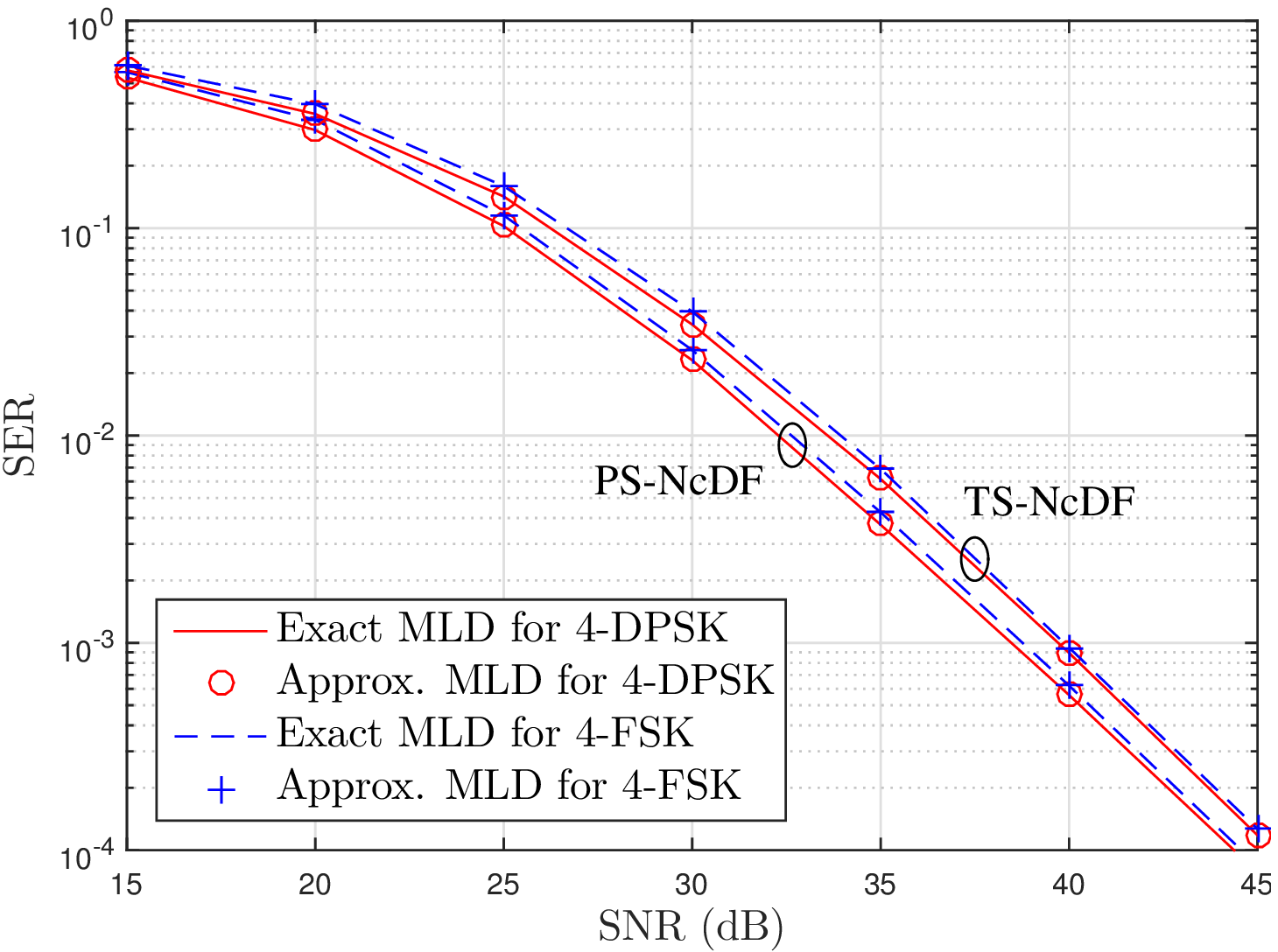}\label{fig:PS_vs_TS_M4_fixed_Ns_SNR}}\\
\subfigure[][$M=8$]
{\includegraphics[width=0.4\columnwidth]{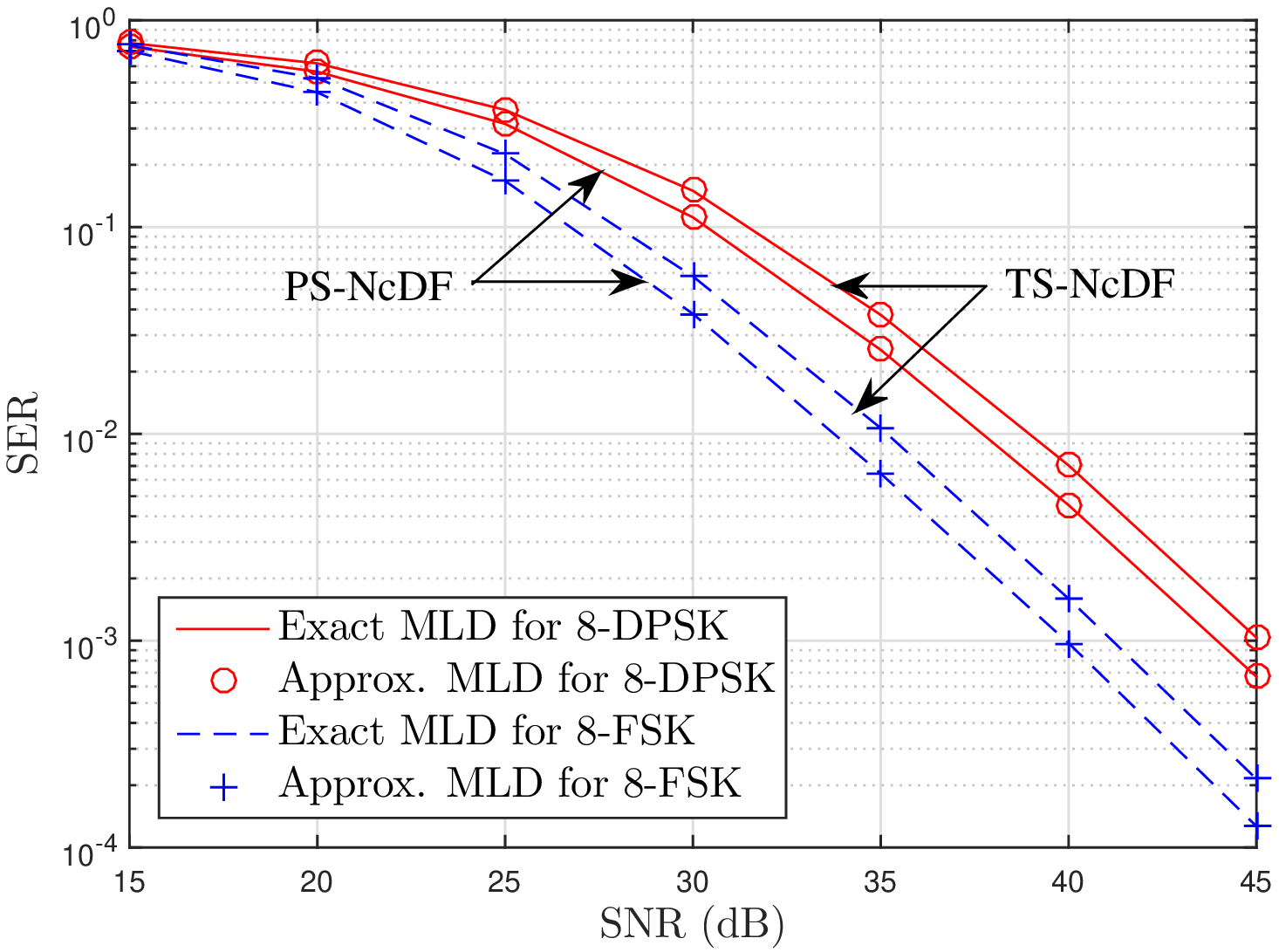}\label{fig:PS_vs_TS_M8_fixed_Ns_SNR}}
\hfil
\subfigure[][$M=16$]
{\includegraphics[width=0.4\columnwidth]{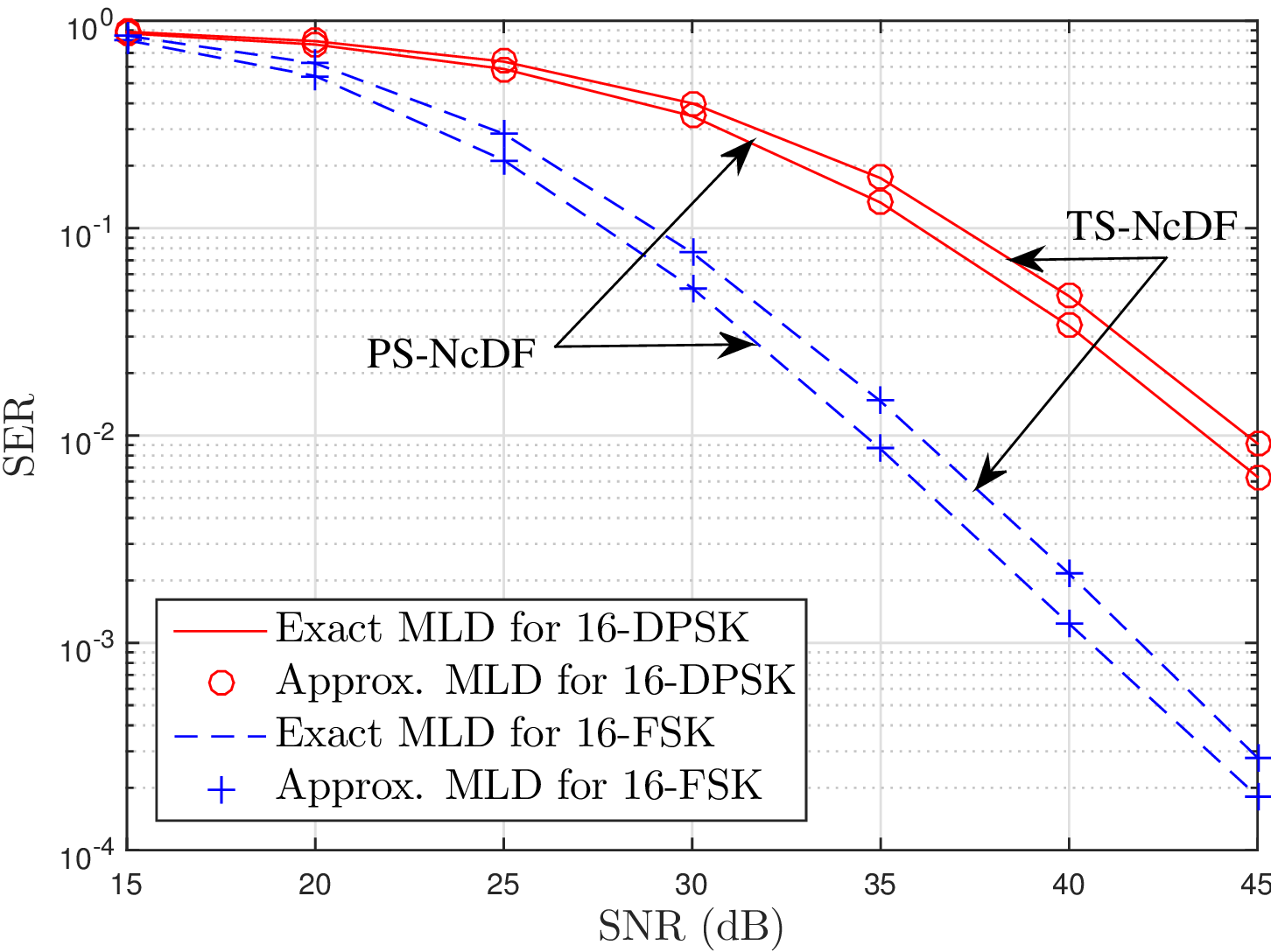}\label{fig:PS_vs_TS_M16_fixed_Ns_SNR}}
\caption{SERs of PS-NcDF and TS-NcDF employing near-optimal EH parameters with $\rho=0.8$ and $\alpha=0.4$ for a single relay network ($K=1$) with $D_{0r}=1.5$ (m): \subref{fig:PS_vs_TS_M2_fixed_Ns_SNR} $M=2$; \subref{fig:PS_vs_TS_M4_fixed_Ns_SNR} $M=4$;  \subref{fig:PS_vs_TS_M8_fixed_Ns_SNR} $M=8$; and  \subref{fig:PS_vs_TS_M16_fixed_Ns_SNR} $M=16$.
}
\label{fig:PS_TS_vary_M}
\end{figure}

\subsection{{\blue Impact of the Modulation Alphabet Size $M$}}
{\blue In this subsection, we are interested in evaluating the performance of a variety of noncoherent signalings for $M = 2, 4, 8, 16$. To that end, we focus on a single-relay network with $D_{0r}=1.5$ (m) employing the near-optimum EH parameters, i.e., $\rho=0.8$ for PS-NcDF and $\alpha=0.4$ for TS-NcDF. Note that $\rho=0.8$ and $\alpha=0.4$ minimize, respectively, the SERs of PS-NcDF and TS-NcDF, as shown in Fig. \ref{fig:PS_TS_fixed_Ns}. However, it is difficult to \emph{analytically} determine the optimum EH parameters. An interesting phenomenon observed is that while $M$-DPSK outperforms $M$-FSK for $M\leq 4$, the performance gain diminishes as $M$ increases, as shown in Figs. \ref{fig:PS_vs_TS_M2_fixed_Ns_SNR} and \ref{fig:PS_vs_TS_M4_fixed_Ns_SNR}. In particular, for $M=4$ shown in Fig. \ref{fig:PS_vs_TS_M4_fixed_Ns_SNR}, the performance gain of DPSK over FSK is almost negligible. As $M$ increases to $M\geq 8$, we see that $M$-FSK achieves a significant gain over $M$-DPSK. In particular, for a target SER of $10^{-2}$, the performance gain of $M$-FSK over $M$-DPSK is approximately 4 dB and 8 dB  for $M=8$ and $16$, respectively, as shown in Figs. \ref{fig:PS_vs_TS_M8_fixed_Ns_SNR} and \ref{fig:PS_vs_TS_M16_fixed_Ns_SNR}. This implies that $M$-FSK achieves a promising energy gain over $M$-DPSK for large $M$.} However, it is important to note that this energy gain comes at the cost of higher required bandwidth. In particular, the minimum double-sideband (DSB) bandwidth at an intermediate frequency (IF) assuming ideal rectangular Nyquist filtering \cite{B.Sklar2001book} is $W=1/T_s$ for $M$-DPSK and $W=M/T_s$ for $M$-FSK ($M$ times that of $M$-DPSK). Nevertheless, considering the significant energy gain, $M$-FSK may be a more suitable solution, especially when $M\geq 8$, for EH relay systems due to the limited energy.


%


\section{Conclusions}\label{sec:con}
We have proposed a noncoherent SWIPT framework embracing both PS-NcDF and TS-NcDF in a unified form, supporting $M$-FSK and $M$-DPSK. Following this framework, we developed exact noncoherent MLDs for PS-NcDF and TS-NcDF in a unified form, which involves integral evaluations. Nevertheless, the exact MLDs are useful for characterizing the optimum performance benchmark for noncoherent SWIPT. Furthermore, to make the MLDs suitable for practical implementation, we developed \emph{closed-form} approximate MLDs achieving near-optimum performance. Numerical results demonstrated a performance tradeoff between the first and second hops through the adjustment of the time switching coefficient or power splitting factors. Moreover, the optimal values of these system parameters corresponding to the minimum SER are strictly between 0 and 1. Finally, we demonstrated that $M$-FSK is an energy-efficient solution for noncoherent SWIPT, especially when $M\geq 8$.

\setcounter{section}{1}

\renewcommand{\thesection}{\Alph{section}}
\renewcommand{\theequation}{\Alph{section}.\arabic{equation}}
\setcounter{equation}{0}
\section*{APPENDIX A: Proof of Lemma \ref{lemma:PDF}}\label{app:PDF}

The conditional distribution of $\bm{X}_0$ given $X_2=z$ is $\bm{X}_0|_{X_2=z}\sim\mathcal{CN}(\bm{0},\bm{\Sigma}_0)$, where
\begin{equation}
  \bm{\Sigma}_0
  =\left[\begin{array}{cc}
     \Omega_1|z|^2+\Omega_0, & \Omega_1|z|^2c^* \\
     \Omega_1|z|^2c, & \Omega_1|z|^2|c|^2+\Omega_0
   \end{array}
    \right].
\end{equation}
It can be shown that $\det(\bm{\Sigma}_0)=\Omega_0\Omega_1|z|^2(1+|c|^2)+\Omega_0^2$ and
\begin{equation}
  \bm{\Sigma}_0^{-1}=\frac{1}{\det(\mathrm{\Sigma_0})}\left[\begin{array}{cc}
     \Omega_1|z|^2|c|^2+\Omega_0, & -\Omega_1|z|^2c^* \\
     -\Omega_1|z|^2c, & \Omega_1|z|^2+\Omega_0
   \end{array}
    \right].
\end{equation}
After some manipulations, the conditional PDF is written as
  \begin{align}
    f_{\bm{X}_0|{X_2=z}}(\bm{x}) = \frac{\exp\left(-\frac{|x_2-cx_1|^2}{\Omega_0(1+|c|^2)}\right)}{\pi^2\Omega_0^2}\frac{\exp\left(-\frac{\frac{|x_1+c^*x_2|^2}{\Omega_0(1+|c|^2)}}
    {1+\frac{\Omega_1}{\Omega_0}(1+|c|^2)|z|^2}\right)}
    {1+\frac{\Omega_1}{\Omega_0}(1+|c|^2)|z|^2}.
 \end{align}
The unconditional PDF is  $f_{\bm{X}_0}(\bm{x})
  =\int_{x=0}^\infty f_{\bm{X}_0\big|{|X_2|^2=x}}(\bm{x})f_{|X_2|^2}(x)\diff x$, which is evaluated to
  \begin{align}
  f_{\bm{X}_0}(\bm{x})
  &=\frac{\e^{-\frac{|x_2-cx_1|^2}{\Omega_0(1+|c|^2)}}}{\pi^2\Omega_0^2} \int_0^\infty \frac{\e^{-\frac{\frac{|x_1+c^*x_2|^2}{\Omega_0(1+|c|^2)}}
    {1+\frac{\Omega_1\Omega_2}{\Omega_0}(1+|c|^2)t}}}
    {1+\frac{\Omega_1\Omega_2}{\Omega_0}(1+|c|^2)t} \e^{-t} \diff t. \label{eq:app_uncon_PDF_DPSK3}
 \end{align}
Applying change of variable $x\leftarrow\frac{\frac{|x_1+c^*x_2|^2}{\Omega_0(1+|c|^2)}}{1+\frac{\Omega_1\Omega_2}{\Omega_0}(1+|c|^2)t}$ in \eqref{eq:app_uncon_PDF_DPSK3} and after some manipulations, we have
  \begin{align}
  f_{\bm{X}_0}(\bm{x})&=\frac{\e^{-\frac{|x_2-cx_1|^2}{\Omega_0(1+|c|^2)}}}{\pi^2\Omega_0^2} \frac{\e^\frac{\Omega_0}{\Omega_1\Omega_2(1+|c|^2)}}{\frac{\Omega_1\Omega_2}{\Omega_0}(1+|c|^2)} \int_0^{\frac{|x_1+c^*x_2|^2}{\Omega_0(1+|c|^2)}} \frac{1}{x} \e^{\Big(x+\frac{|x_1+c^*x_2|}{\Omega_1\Omega_2(1+|c|^2)^2x}\Big)}\diff x, \label{eq:app_uncon_PDF_DPSK4}
 \end{align}
 which is the same as \eqref{eq:dist_X0} upon some appropriate parameterizations.

The derivation of the PDF of $\bm{Y}_0$ follows similar lines as the derivation of $f_{\bm{X}_0}(\bm{x})$, which is briefly summarized as follows. The conditional PDF of $\bm{Y}_0$ given $|X_2|^2=x$ is given by
\begin{align}
  f_{\bm{Y}_0\big||X_2|^2=x}(\bm{y})&=\frac{1}{(\pi\Omega_0)^M}\exp\Bigg(-\frac{1}{\Omega_0}\sum_{\substack{q=1\\ q\neq p}}^M|y_q|^2\Bigg)\frac{1}{1+\frac{\Omega_1}{\Omega_0}x}\exp\bigg(-\frac{|y_p|^2}{\Omega_0+\Omega_1x}\bigg).
\end{align}
Similarly, the unconditional PDF is obtained as
\begin{align}
  f_{\bm{Y}_0}(\bm{y})
  &=\frac{1}{(\pi\Omega_0)^M}\e^{-\frac{\|\bm{y}\|^2-|y_p|^2}{\Omega_0}}\int_0^\infty \frac{1}{1+\frac{\Omega_1}{\Omega_0}x}\exp\bigg(-\frac{|y_p|^2}{\Omega_0+\Omega_1x}\bigg)\frac{1}{\Omega_2}\e^{-\frac{x}{\Omega_2}}\diff x. \label{eq:app_uncon_PDF_FSK}
\end{align}
Using similar change of variable in \eqref{eq:app_uncon_PDF_FSK}, we can obtain the final expression of \eqref{eq:dist_Y0}.

\setcounter{section}{2}

\renewcommand{\thesection}{\Alph{section}}
\renewcommand{\theequation}{\Alph{section}.\arabic{equation}}
\setcounter{equation}{0}
\section*{APPENDIX B: Proof of Lemma \ref{lemma:phase_dist}}\label{app:phase_dist}
It follows that $\bm{Y}_2 \sim\mathcal{CN}(\bm{0},\bm{\Sigma}_2)$, where
\begin{align}
  \bm{\Sigma}_2=\Omega_0 \left[
                            \begin{array}{cc}
                              1+\gamma, & \gamma c^* \\
                              \gamma c, & 1+\gamma|c|^2 \\
                            \end{array}
                          \right],
\end{align}
where $\gamma = \Omega_1/\Omega_0$. Applying the property of the multivariate Gaussian distribution, the conditional distribution of $Y_2$ given $Y_1$ follows a Gaussian distribution, i.e., $Y_2|_{Y_1=y_1}\sim\mathcal{CN}(\vartheta_0y_1,\sigma_0^2)$, where $\vartheta_0 \pardef\frac{\gamma c}{1+\gamma}$ and $\sigma_0^2 \pardef \Omega_0\left(1+\frac{\gamma}{1+\gamma}|c|^2\right)$.
Let $Z_0\pardef Y_1^*Y_2$. It follows that $Z_0|_{Y_1=y_1}\sim\mathcal{CN}(\vartheta_0|y_1|^2,\sigma_0^2|y_1|^2)$. Thus, the unconditional PDF of $Z_0$ is
 \begin{align}
  f_{Z_0}(z)
  &=\frac{\exp\Big(\frac{2\Re\{\vartheta_0^*z\}}{\sigma_0^2}\Big)}{\pi\sigma_0^2(\Omega_0+\Omega_1)}\int_0^\infty \frac{1}{x}\exp \bigg\{-\bigg[\bigg(\frac{1}{\Omega_0+\Omega_1}+\frac{|\vartheta_0|^2}{\sigma_0^2}\bigg)x+\frac{|z|^2}{\sigma_0^2x}\bigg]\bigg\}\diff x. \label{eq:app_Z_pdf1}
\end{align}
According to \cite[eq. (3.471.9)]{I.S.Gradshteyn2007book}, \eqref{eq:app_Z_pdf1} is evaluated to
\begin{align}
  f_{Z_0}(z)&=\frac{2}{\pi\sigma_0^2(\Omega_0+\Omega_1)}\exp\bigg(\frac{2\Re\{\vartheta_0^*z\}}{\sigma_0^2}\bigg)K_0\Bigg(2\sqrt{\frac{1}{\Omega_0+\Omega_1}+
  \frac{|\vartheta_0|^2}{\sigma_0^2}}\frac{|z|}{\sigma_0}\Bigg), \label{eq:app_Z_pdf1}
\end{align}
where $K_0(x)$ is the zero-order modified Bessel function of the second kind \cite[eq. (8.432.1)]{I.S.Gradshteyn2007book}. Using the cartesian-to-polar transformation, $R_0\e^{j\Theta_0}\pardef Z_0$, the joint PDF of $R_0$ and $\Theta_0$ is
\begin{align}
\!\!\!\!\!\!\!\!  f_{R_0,\Theta_0}(r_0,\theta_0) = \frac{2}{\pi\sigma_0^2(\Omega_0+\Omega_1)}r_0\exp\bigg[\frac{2|\vartheta_0|r_0}{\sigma_0^2}\cos(\theta_0-\measuredangle c)\bigg]K_0\Bigg(\frac{2r_0}{\sigma_0}\sqrt{\frac{1}{\Omega_0+\Omega_1}+
  \frac{|\vartheta_0|^2}{\sigma_0^2}}\Bigg). \!\!
\end{align}
Since $\Theta=\Theta_0-\measuredangle c$, the PDF of $\Theta$ is $f_{\Theta}(\theta) = \int_0^\infty f_{R_0,\Theta_0}(r_0,\theta + \measuredangle c) \diff r_0$, which evaluates to
\begin{align}
  f_{\Theta}(\theta)
   & = \frac{2}{\pi\sigma_0^2(\Omega_0+\Omega_1)}\int_0^\infty r_0\exp\bigg(\frac{2|\vartheta_0|r_0}{\sigma_0^2}\cos\theta\bigg)K_0\Bigg(\frac{2r_0}{\sigma_0}\sqrt{\frac{1}{\Omega_0+\Omega_1}+
  \frac{|\vartheta_0|^2}{\sigma_0^2}}\Bigg) \diff r_0.\label{eq:app_theta_pdf1}
\end{align}
It is verified by Mathematica$^\circledR$ that the following equation holds\footnote{An alternative closed-form solution of the integral in \eqref{eq:app_integral} is available in \cite[eq. (6.621.3)]{I.S.Gradshteyn2007book}. However, that expression is given in terms of the Gaussian hypergeometric function, which is more complex than the expression in \eqref{eq:app_integral}.}
\begin{align}
  \int_0^\infty x\e^{-a x}K_0(bx)dx=\frac{1}{b^2-a^2}-\frac{a\arccos\frac{a}{b}}{(b^2-a^2)^{3/2}}\label{eq:app_integral}
\end{align}
for $b>0$ and $\Re(a+b)>0$. Letting $a=-\frac{2|\vartheta_0|r_0}{\sigma_0^2}\cos\theta$ and $b=\frac{2}{\sigma_0}\sqrt{\frac{1}{\Omega_0+\Omega_1}+
  \frac{|\vartheta_0|^2}{\sigma_0^2}}$ in \eqref{eq:app_integral}, we solve the integral of \eqref{eq:app_theta_pdf1} in closed-form as
\begin{align*}
  f_{\Theta}(\theta) &= \frac{1}{2\pi\Big[1+\frac{|\vartheta_0|^2(\Omega_0+\Omega_1)}{\sigma_0^2}\sin^2\theta\Big]}\Bigg[1+\frac{\cos\theta}{\sqrt{\sin^2\theta+
  \frac{\sigma_0^2}{|\vartheta_0|^2(\Omega_0+\Omega_1)}}}\arccos\Bigg(-\frac{\cos\theta}{\sqrt{\sin^2\theta+
  \frac{\sigma_0^2}{|\vartheta_0|^2(\Omega_0+\Omega_1)}}}\Bigg)\Bigg].
\end{align*}
Finally, substituting $\vartheta_0$ and $\sigma_0^2$ into the above expression yields the final expression in \eqref{eq:phase_dist}.

\setcounter{section}{3}

\renewcommand{\thesection}{\Alph{section}}
\renewcommand{\theequation}{\Alph{section}.\arabic{equation}}
\setcounter{equation}{0}
\section*{APPENDIX C: Proof of Lemma \ref{lemma:numerical_computation}}\label{app:numerical_computation}
From \eqref{eq:alternative2}, we can rewrite $I(\epsilon,\beta)$ for $\beta\epsilon>1$ as
  \begin{align}
    I(\epsilon,\beta)=\frac{\e^{\frac{1}{\epsilon}}}{\epsilon}K_0\bigg(2\sqrt{\frac{\beta}{\epsilon}}\bigg)+ \int_{\frac{1}{\epsilon}}^{\sqrt{\frac{\beta}{\epsilon}}}\frac{1}{\epsilon x}\e^{-\big(x+\frac{1}{\epsilon}\big)}\e^{-\frac{\beta}{\epsilon x}}\diff x.\label{eq:app_I1}
  \end{align}
Over the integration interval $(1/\epsilon, \sqrt{\beta/\epsilon})$ in \eqref{eq:app_I1}, we have $0<\frac{1}{\epsilon x}<1$, $0<\e^{-(x+\frac{1}{\epsilon})}<1$, and $0<\e^{-\frac{\beta}{\epsilon x}}<1$, meaning that the overall integrand is a finite bounded function. Moreover, the integrand is always between 0 and 1, which ensures fast convergence when approximated by the Gauss-Legendre quadrature (GLQ) \cite{P.J.Davis1984book}. Note that the GLQ is one of the most efficient and accurate non-adaptive quadrature rules for a given number of arithmetic operations \cite{P.J.Davis1984book}. Applying the $N$-th order GLQ rule of \cite[eq. (25.4.29)]{M.Abramowitz1964book} in \eqref{eq:app_I1} yields $I(\epsilon,\beta)=I_N(\epsilon,\beta)+R_N$, where $I_N(\epsilon,\beta) = \frac{\e^{\frac{1}{\epsilon}}}{\epsilon}\big[K_0\big(2\sqrt{\frac{\beta}{\epsilon}}\big)+\Psi_N\big(\frac{1}{\epsilon},\sqrt{\frac{\beta}{\epsilon}}\big)
        \big]$
and $\Psi_N(\cdot,\cdot)$ is given in \eqref{eq:Psi}. This gives the third case of \eqref{eq:N_approx}. For $\beta\epsilon<1$ in \eqref{eq:alternative2},
it is easy to verify that the overall integrand is not necessarily smaller than one in this case. Thus, simply applying the GLQ rule for $\beta\epsilon<1$ in \eqref{eq:alternative2} might give an approximation with slower convergence speed compared to the GLQ approximation for $\beta\epsilon>1$. However, if we further enforce $\frac{\e^{\frac{1}{\epsilon}}}{\epsilon}\leq 1$, then the pre-bracket coefficient $\frac{\e^{\frac{1}{\epsilon}}}{\epsilon}$ in \eqref{eq:alternative2} can scale down the approximation error for the integral, which can significantly accelerate the convergence speed for the approximation. Thus, for $\beta\epsilon<1$ and $\frac{\e^{\frac{1}{\epsilon}}}{\epsilon}\leq 1$, the $N$-th order GLQ rule of \cite[eq. (25.4.29)]{M.Abramowitz1964book} readily applies, which yields $  I_N(\epsilon,\beta) = \frac{\e^{\frac{1}{\epsilon}}}{\epsilon}\big[K_0\big(2\sqrt{\frac{\beta}{\epsilon}}\big)-\Psi_N\big(\beta,\sqrt{\frac{\beta}{\epsilon}}\big)
        \big]$.
This is the same as the second case of \eqref{eq:N_approx}. For the remaining case with $\beta\epsilon<1$ and $\frac{\e^{\frac{1}{\epsilon}}}{\epsilon}>1$, we start with the alternative form of $I(\epsilon,\beta)$ in \eqref{eq:alternative1}, which can be written as $I(\epsilon,\beta)=\mathbb{E}\big[\frac{1}{1+\epsilon X_0}\exp\big(-\frac{\beta}{1+\epsilon X_0}\big)\big]$
where the expectation is taken with respect to $X_0$ and $X_0$ is an exponential random variable with unit mean. Thus, the integral $I(\epsilon,\beta)$ can be interpreted as the statistical mean of the random variable $g(X_0)\pardef \frac{1}{1+\epsilon X_0}\exp\big(-\frac{\beta}{1+\epsilon X_0}\big)$. Using the $N$-th order estimate of the mean of random variables, we can approximate $I(\epsilon,\beta)$ by $I_N(\epsilon,\beta)$ as follows \cite[eq. (5.85)]{A.Papoulis2002book}:
\begin{align}
  I_N(\epsilon,\beta)=\sum_{n=0}^N\frac{\mu_n}{n!}\frac{\diff^n g(x)}{\diff x}\Bigg|_{x=E[X_0]=1},\label{eq:app_talor1}
\end{align}
where $\mu_n=\mathbb{E}[|X_0-E[X_0]|^n]$, $n=0,\cdots,N$. Using the Binomial theorem and the exponential distribution of $X_0$, it is not hard to show that $\mu_n=\sum_{i=0}^n\frac{(-1)^i}{i!}n!$. The function $g(x)$ can be represented as a composite function with $g(x)=f_0[g_0(x)]$, where $f_0(t)\pardef t\e^{-\beta t}$ and $t=g_0(x)\pardef1/(1+\epsilon x)$. Thus, Fa$\grave{\text{a}}$ di Bruno's formula \cite{W.P.Johnson2002.3} is applied as follows:
\begin{align}
  \frac{\diff^n f_0[g_0(x)]}{\diff x}\Bigg|_{x=1} = \sum\limits_{\substack{\sum_{i=1}^n im_i=n\\  \forall m_i\in \mathbb{Z}^{+}}}\frac{n!}{\prod\limits_{j=1}^n m_j!}f_0^{(\varsigma)}(t)\bigg|_{t=\tau}\prod_{k=1}^n\bigg(\frac{g_0^{(k)}(x)}{k!}\bigg)^{m_k}\bigg |_{x=1},\label{eq:app_taylor_der}
\end{align}
where $\tau = g_0(1)$, $\varsigma=\sum_{i=1}^n m_i$, and $\mathbb{Z}^{+}\pardef\{0,1,2,\cdots\}$. It can be shown that $f_0^{(\varsigma)}(t)=(\varsigma-\beta t)(-\beta)^{\varsigma-1}\e^{-\beta t}$ and $g_0^{(k)}(x)=\frac{(-1)^k k! \epsilon^k}{(\epsilon x + 1)^{k+1}}$. Applying these relationships in \eqref{eq:app_taylor_der}, substituting the expression into \eqref{eq:app_talor1}, and after tedious manipulations,  we obtain the first case of \eqref{eq:N_approx}.

\setcounter{section}{4}

\renewcommand{\thesection}{\Alph{section}}
\renewcommand{\theequation}{\Alph{section}.\arabic{equation}}
\setcounter{equation}{0}
\section*{APPENDIX D: Proof of Lemma \ref{lemma:transition_prob}}\label{app:transition_prob}

According to the MLD rule in \eqref{eq:MLD_relay_DPSK}, the phase $\Theta_r\pardef \measuredangle y_{0r-1}^*(l)y_{0r}(l)$ is a sufficient statistic for the MLD. Let $\mathcal{D}_m\pardef \{\theta| \theta_m-\frac{\pi}{M}\leq \theta \leq \theta_m+\frac{\pi}{M}\}$ denote the decision region for the message $m$, where $\theta_m\pardef \frac{2\pi m}{M}$, $m=0,\cdots,M-1$. Then the ML decision rule is in favor of the message $m_r$ if $\Theta_r\in \mathcal{D}_{m_r}$, $m_r=0,\cdots,M-1$. Thus, the transition probability is written as
\begin{align}
  \Pr(m_r|m)
  &=\Pr\bigg(\frac{\big[2(m_r-m)-1\big]\pi}{M}\leq \Theta_r-\theta_m \leq \frac{\big[2(m_r-m)+1\big]\pi}{M}\bigg|m\bigg).\label{eq:app_TP2}
\end{align}

The unified system model for $\mathrm{T}_r$ in \eqref{eq:y0r_DPSK_unified} can be rewritten as $\bm{y}_{0r}=\hbar_{0r}[1,\e^{j\theta_m}]^T+\bm{n}^\prime_{0r}$, where $\hbar_{0r}\pardef \sigma_{0r}\sqrt{\gamma_{0r}}h_{0r}s(l-1)$ and $\bm{n}^\prime_{0r}\pardef \sigma_{0r}\bm{n}_{0r}$. It follows that $\hbar_{0r}\sim\mathcal{CN}(0,\sigma_{0r}^2\gamma_{0r})$ and $\bm{n}^\prime_{0r}\sim\mathcal{CN}(0,\sigma_{0r}^2)$. Applying the result in Lemma \ref{lemma:phase_dist}, the PDF of $\Theta^\prime_r\pardef \Theta_r-\theta_m$ is given by
  \begin{align}
    f_{\Theta^\prime_r}(\theta)=\frac{1}{2\pi}\frac{1}{1+\frac{\gamma_{0r}^2\sin^2\theta}{1+2\gamma_{0r}}}\Bigg [1+\frac{\gamma_{0r}\cos\theta}{\sqrt{\gamma_{0r}^2\sin^2\theta+2\gamma_{0r}+1}}\arccos\Bigg(-\frac{\gamma_{0r}\cos\theta}{\gamma_{0r}+1}\Bigg)\Bigg].
    \label{eq:app_phase_dist}
  \end{align}
Thus, the transition probability in \eqref{eq:app_TP2} can be written as
\begin{subequations}\label{eq:app_symmetry_TP}
\begin{align}
  \Pr(m_r|m)&=\int_{\frac{2(m_r-m)-1}{M}\pi}^{\frac{2(m_r-m)+1}{M}\pi}f_{\Theta^\prime_r}(\theta)\diff \theta\label{eq:app_TP3}\\
  &=\int_{\frac{2(m-m_r)-1}{M}\pi}^{\frac{2(m-m_r)+1}{M}\pi}f_{\Theta^\prime_r}(\alpha)\diff \alpha\label{eq:app_TP5}\\
  &=\Pr(m|m_r),\label{eq:app_TP6}
\end{align}
\end{subequations}
where \eqref{eq:app_TP5} is due to the change of variable $\alpha\leftarrow -\theta$ and $f_{\Theta^\prime_r}(\theta)=f_{\Theta^\prime_r}(-\theta)$. Note that the symmetry $\Pr(m_r|m)=\Pr(m|m_r)$ can be also attributed to the rotational invariance property of the PSK signal constellation. According to \eqref{eq:app_symmetry_TP}, the transition probability can be calculated as
\begin{align}
  \Pr(m_r|m)&=\int_{\frac{2|m_r-m|-1}{M}\pi}^{\frac{2|m_r-m|+1}{M}\pi}f_{\Theta^\prime_r}(\theta)\diff \theta.\label{eq:app_TP7}
\end{align}
Finally, substituting \eqref{eq:app_phase_dist} into \eqref{eq:app_TP7} yields the final expression in \eqref{eq:transition_prob}.

\setcounter{section}{5}

\renewcommand{\thesection}{\Alph{section}}
\renewcommand{\theequation}{\Alph{section}.\arabic{equation}}
\setcounter{equation}{0}
\section*{APPENDIX E: Proof of Lemma \ref{lemma:transition_prob_approx}}\label{app:transition_prob_approx}
Noticing that the transmitted PSK phase will most likely be mistaken for its two closest neighboring phases with equal probability \cite{J.G.Proakis2000book}, we can approximate the transition probability as follows: for the two closest error transitions with $|m_r-m|=1$ or $M-1$, we have $\Pr(m_r|m)\approx \frac{\mathrm{SER}}{2}$; For the correct symbol transition with $m_r=m$, we have $\Pr(m_r|m)=1-\mathrm{SER}$. For all other error transitions, we simply ignore the probabilities. This approximation immediately gives \eqref{eq:TP_approx_by_SER}. It remains to derive the SER for $M$-DPSK. For $M=2$, a closed-from exact expression of the SER is available, namely, $\mathrm{SER}=\varepsilon_r^{\rm p} = \frac{1}{2(1+\gamma_{0r})}$. For $M\geq 4$, however, no closed-form exact SER expression is reported in the literature for fading channels. To proceed, we resort to an upper bound on the SER for AWGN channels in \cite[eq. (8.95)]{M.K.Simon2005book}, which is the best known bound for AWGN channels. Taking the statistical average of this bound over the Rayleigh distribution by means of \cite[eq. (6.283.2)]{I.S.Gradshteyn2007book}, we obtain a novel upper bound on the (average) SER of $M$-DPSK as $\varepsilon_r^{\rm p}    =1.03\sqrt{\frac{1+\cos\frac{\pi}{M}}{2\cos\frac{\pi}{M}}}\big[1-\sqrt{\frac{(1-\cos\frac{\pi}{M})\gamma_{0r}}{1+(1-\cos\frac{\pi}{M})
    \gamma_{0r}}}\big]$.
Overall, the SER of $M$-DPSK is given by \eqref{eq:PSK_relay_SER}.


\setcounter{section}{6}

\renewcommand{\thesection}{\Alph{section}}
\renewcommand{\theequation}{\Alph{section}.\arabic{equation}}
\setcounter{equation}{0}
\section*{APPENDIX F: Proof of Theorem \ref{thm:Exact_MLD}}\label{app:Exact_MLD}
Applying Total Probability Theorem and Lemma \ref{lemma:transition_prob},  the likelihood function (LF) conditioned on message $m$ for the unified $M$-DPSK model in \eqref{eq:DPSK_unified} can be written as
\begin{align}
  f(\bm{y}_{0d},\{\bm{y}_{rd}\}_{r=1}^K|m)
  &=f(\bm{y}_{0d}|m)\prod_{r=1}^K \sum_{m_r=0}^{M-1}f(\bm{y}_{rd}|m_r)\mathcal{P}_{|m_r-m|}(\gamma_{0r}).\label{eq:app_PSK_LF3}
\end{align}
From \eqref{eq:y0d_DPSK_unified}, we see that $\bm{y}_{0d}|_m\sim\mathcal{CN}(\bm{0},\bm{\Sigma}_{0d})$, where
\begin{align}
  \bm{\Sigma}_{0d}=\sigma_{0d}^2\left[
                             \begin{array}{cc}
                               1+\gamma_{0d} & \gamma_{0d}\e^{-j2\pi m/M} \\
                               \gamma_{0d}\e^{j2\pi m/M} & 1+\gamma_{0d} \\
                             \end{array}
                           \right].
\end{align}
After some manipulations, we have
  \begin{align}
    f(\bm{y}_{0d}|m) = \frac{\exp\left(-\frac{(1+\gamma_{0d})\|\bm{y}_{0d}\|^2}{\sigma_{0d}^2(1+2\gamma_{0d})}\right)}{(\pi\sigma_{0d}^2)^2(1+2\gamma_{0d})}\exp\bigg\{
     \frac{2\gamma_{0d}}{1+2\gamma_{0d}}\frac{\Re\{y_{0d}(l-1)y_{0d}^*(l)\e^{j2\pi m/M}\}}{\sigma_{0d}^2}\bigg\}. \label{eq:app_LF_0d_PSK}
 \end{align}
From \eqref{eq:yrd_DPSK_unified}, we see that applying Lemma \ref{lemma:PDF} immediately yields the LF $f(\bm{y}_{rd}|m_r)$ as follows:
  \begin{align}
\!\!\!    f(\bm{y}_{rd}|m_r)= \frac{\exp\bigg(-\frac{\big|y_{rd}(l)-y_{rd}(l-1)\e^{j2\pi m_r/M}\big|^2}{2\sigma_{rd}^2}\bigg)}{(\pi\sigma_{rd}^2)^2}I\bigg(2\gamma_{rd}, \frac{\big|y_{rd}(l)+y_{rd}(l-1)\e^{j2\pi m_r/M}\big|^2}{2\sigma_{rd}^2}\bigg).\label{eq:app_LF_rd_PSK}
  \end{align}
Substituting \eqref{eq:app_LF_0d_PSK} and \eqref{eq:app_LF_rd_PSK} into \eqref{eq:app_PSK_LF3} and using the definitions of $\beta_0(m)$, $\beta_r^{+}(m_r)$, and $\beta_r^{-}(m_r)$ given in Theorem  \ref{thm:Exact_MLD}, we obtain the MLD in \eqref{eq:MLD_DPSK} for $M$-DPSK. For $M$-FSK in \eqref{eq:FSK_unified}, following similar procedure as in \eqref{eq:app_PSK_LF3} for $M$-DPSK,  the LF can be written as
\begin{align}
\!\!\! \!\!\! f(\by_{0d},\{\by_{rd}\}_{r=1}^K|m)
  &\!=\!f(\by_{0d}|m)\prod_{r=1}^K \! \sum_{m_r=0}^{M-1}\!\bigg[(1-\epsilon_r)f(\by_{rd}|m_r=m)\!+\!\frac{\epsilon_r}{M-1}\!\!\!\sum_{m_r\neq m}f(\by_{rd}|m_r)\bigg],\label{eq:app_FSK_LF2}
\end{align}
Furthermore,  the PDF of $\by_{0d}$ in \eqref{eq:y0d_FSK_unified}, conditioned on the knowledge of $m$, is
\begin{align}
  f(\by_{0d}|m)=\frac{\e^{-\frac{\|\by_{0d}\|^2}{\sigma_{0d}^2}}}{(\pi\sigma_{0d}^2)^M(1+\gamma_{0d})}\e^{\frac{\gamma_{0d}}{1+\gamma_{0d}}
  \frac{|\mathrm{y}_{0d}(m+1)|^2}{\sigma_{0d}^2}}.\label{eq:app_cond_LF_0d_FSK}
\end{align}
Applying Lemma \ref{lemma:PDF}, the PDF of $\by_{rd}$ in \eqref{eq:yrd_FSK_unified} given $m_r$ is given by
\begin{align}
  f(\by_{rd}|m_r)=\frac{1}{(\pi\sigma_{rd}^2)^M}\e^{-\frac{\|\by_{rd}\|^2-|\mathrm{y}_{rd}(m_r+1)|^2}{\sigma_{rd}^2}} I\bigg(\gamma_{rd},\frac{|\mathrm{y}_{rd}(m_r+1)|^2}{\sigma_{rd}^2}\bigg).\label{eq:app_cond_LF_rd_FSK}
\end{align}
Finally, substituting \eqref{eq:app_cond_LF_0d_FSK} and \eqref{eq:app_cond_LF_rd_FSK} into \eqref{eq:app_FSK_LF2} yields the MLD in \eqref{eq:MLD_FSK} for $M$-FSK.

\end{document}